\documentclass[aps,prd,nofootinbib,showpacs,preprintnumbers,longbibliography,amssymb,11pt]{revtex4-1} 
\usepackage[sort&compress]{natbib}
\usepackage[dvipsnames]{xcolor}

\usepackage{graphicx}
\usepackage{epsfig}
\usepackage{dcolumn}
\usepackage{bm}

\usepackage{amsmath}
\usepackage{siunitx}
\usepackage{amssymb}
\usepackage{physics}
\usepackage{slashed}
\usepackage{dsfont}
\usepackage{subcaption}
\usepackage{hyperref}

\graphicspath{{./Figures/}}

\newcommand{\eqnref}[1]{Eq.~(\ref{eqn:#1})}
\newcommand{\eqnsref}[2]{Eqs.~(\ref{eqn:#1}) and (\ref{eqn:#2})}
\newcommand{\secref}[1]{Sec.~\ref{sec:#1}}

\newcommand{\subsecref}[1]{Subsec.~\ref{subsec:#1}}

\newcommand{\figref}[1]{Fig.~\ref{fig:#1}}
\newcommand{\figsref}[2]{Figs.~\ref{fig:#1} and \ref{fig:#2}}

\newcommand{\tableref}[1]{Table~\ref{table:#1}}

\begin{document}

\preprint{MITP-22-074}

\title{Axion Couplings in Gauged $U(1)'$ Extensions of the Standard Model}

\author{Alexey Kivel}
\email{alkivel@uni-mainz.de}
\author{Julien Laux}
\email{jlaux01@uni-mainz.de}
\author{Felix Yu}
\email{yu001@uni-mainz.de}
\affiliation{PRISMA+ Cluster of Excellence \& Mainz Institute for
  Theoretical Physics, Johannes Gutenberg University, 55099 Mainz,
  Germany}

\begin{abstract}
We explore the effective theory of an axion in a gauged baryon number symmetry extension of the Standard Model (SM), where the axion is realized from a Dine-Fischler-Srednicki-Zhitnitsky (DFSZ) model construction. Integrating out the anomalons realizes a pattern of effective Wilson coefficients reflecting the factorization between the anomalous Peccei-Quinn and gauged baryon number symmetries. We construct and analyze the chiral transformation invariance of the axion effective theory, accounting for possible flavor-violating axion couplings. We calculate and study the  unique phenomenology of the axion and $Z'$ boson, and we present the current collider limits on these particles in the $\{m_{Z'}, g_B\}$, $\{m_a, G_{a \gamma \gamma} \}$, and $\{m_a, f_a^{-1}\}$ planes.
\end{abstract}

\maketitle

\section{Introduction}
\label{sec:Introduction}

The study of fundamental symmetries of Nature is central to the development of the Standard Model as well as its possible extensions.  In particular, the basic structure of the Standard Model (SM) is built from the gauge symmetry groups $SU(3)_c \times SU(2)_L \times U(1)_Y$, while the three generations of SM fermions admit a residual $U(1)_B \times U(1)_L$ global symmetry after accounting for arbitrary Yukawa interactions and massive neutrinos.  While the role of the SM gauge symmetries is well-understood as conserved current interactions among SM matter fields (albeit spontaneously broken in the case of the electroweak symmetry), the global symmetries have no corresponding low-energy mediators and are in fact expected to be violated individually because of their anomalous nature.  In contrast, the Peccei-Quinn symmetry~\cite{Peccei:1977hh, Peccei:1977ur} is also an anomalous global symmetry that is spontaneously broken at a high scale $f_a$, leading to a very light axion degree of freedom~\cite{Weinberg:1977ma, Wilczek:1977pj, Kim:1979if, Shifman:1979if, Zhitnitsky:1980tq, Dine:1981rt} that is responsible for the resolution of the strong CP problem.  Given that the non-observation of the electric dipole moment of the neutron strongly constrains the anomalous theta term $\bar{\theta}G \tilde{G}$ of the SM, with $\bar{\theta} \equiv \theta + \arg \det M_q$, to be $\bar{\theta} \lesssim 10^{-10}$~\cite{Baker:2006ts, Pendlebury:2015lrz, ParticleDataGroup:2020ssz}, experimental searches for axions are highly motivated.

From this viewpoint, we want to consider the interplay between gauged anomalous global symmetries of the SM and the Peccei-Quinn (PQ) symmetry, with a particular focus on the distinction between the corresponding $Z'$ and axion degrees of freedom of the two symmetries, respectively.  For concreteness, we will gauge SM baryon number, which necessarily requires the addition of new electroweak-charged fermions called anomalons to cancel the $SU(2)_L^2 \times U(1)_B$ and $U(1)_Y^2 \times U(1)_B$ anomalies~\cite{Carone:1994aa, Carone:1995pu, FileviezPerez:2010gw, FileviezPerez:2011pt, Duerr:2013dza, Dobrescu:2013cmh, Dobrescu:2014fca, Michaels:2020fzj, Dobrescu:2021vak}.  To connect to axion physics, we will also add two $SU(2)_L$ Higgs doublets and two $U(1)_B$ Higgs fields with Dine-Fischler-Srednicki-Zhitnitsky (DFSZ) model-like couplings~\cite{Zhitnitsky:1980tq, Dine:1981rt}.  In this way, the usual SM gauge singlet in DFSZ axion models is now the remnant pseudo-Nambu Goldstone boson from $U(1)_B$ symmetry breaking and is orthogonal to the Goldstone eaten by the $Z'$ vector boson, enabling us to explore the nontrivial dynamics of these Goldstone fields.

Furthermore, we are focused on the effective description of axions and axion-like particles in the presence of new $U(1)$ gauge symmetries.  While the SM effective field theory (EFT) of axions and axion-like particles (ALPs) has been extensively discussed recently~\cite{Mimasu:2014nea, Jaeckel:2015jla, Knapen:2016moh, Izaguirre:2016dfi, Brivio:2017ije, Bauer:2017nlg, Bauer:2017ris, Bauer:2018uxu, Quevillon:2019zrd, Gavela:2019cmq, Bauer:2019gfk, DiLuzio:2020wdo, Bauer:2020jbp, Galda:2021hbr, Bonilla:2021ufe, Bauer:2021mvw}, the study of anomalous PQ breaking and the corresponding assumptions on the effective field theory are crucially important.  As an example, models featuring nontrivial completions of the quantum chromodynamics (QCD) gauge symmetry can shift the viable axion mass range into the electroweak scale and heavier~\cite{Agrawal:2017ksf, Agrawal:2017evu, Gaillard:2018xgk, Kivel:2022emq}, marking a new regime for axion effective field theory.

Our top-down approach to axion+$Z'$ effective field theory from a gauged baryon number and DFSZ axion model predicts specific patterns of couplings of the axion and $Z'$ degrees of freedom to the SM fermions and gauge bosons.  We will show that these patterns are a direct result of the ultraviolet (UV) requirement of anomaly cancellation, the one-loop trace condition to eliminate cut-off scale dependent kinetic mixing~\cite{Dobrescu:2021vak}, and the requirement for the charges of the new matter fields to allow decays into SM fields in the early universe, avoiding new color- or EM-charged stable relics.  We will also emphasize the new collider observables that result from our work, which include decays such as $a \to hZ$ and $Z' \to a h$ signatures.

The outline of our paper is as follows.  In~\secref{Modelbuilding}, we present the gauged baryon model augmented by the DFSZ-like axion scalar sector.  We will emphasize the role of the anomalon sector in determining the absence of chiral anomalies and log-divergent kinetic mixing, as well as the origin of the Peccei-Quinn symmetry from the scalar potential.  In~\secref{EFT} we present a general calculation of EFT operators in an axion interaction basis focusing on the operators mediating axion-vector-vector couplings and axion-vector-Higgs couplings.  Our calculations are presented in~\secref{WilsonCoefficients} as generic loop functions that allow for flavor-changing vertices of the intermediate fermions.  We discuss the phenomenological aspects of our model in~\secref{pheno} and highlight the new discovery and search channels involving the $Z'$ and axion particles at colliders.
We conclude in~\secref{conclusions}.

\section{DFSZ axion with gauged baryon number}
\label{sec:Modelbuilding}

To analyze the structure of the anomalous Peccei-Quinn symmetry in $U(1)'$-gauge symmetry extensions of the Standard Model, we study a DFSZ-scalar extension to a gauged baryon number model.  Our gauged baryon number model is adapted from Refs.~\cite{Dobrescu:2014fca, Dobrescu:2021vak}, where the new fermion content is chosen to cancel the $SU(2)^2 \times U(1)_B$ and $U(1)_Y^2 \times U(1)_B$ chiral anomalies arising from the SM quarks.  Furthermore, the $U(1)_B$ charges of these anomalons also satisfy the trace condition that naturally suppresses kinetic mixing between the $Z'$ and the SM $Z$ and photon~\cite{Dobrescu:2021vak}.

The field content is shown in~\tableref{U1B_DFSZ}, where the anomalon fields are $L'_L$, $L'_R$, $E'_L$, $E'_R$, $N'_L$, and $N'_R$, which have the indicated SM and $U(1)_B$ gauge charges designed to cancel the $U(1)_B$ gauge anomalies and satisfy the trace orthogonality condition for kinetic mixing~\cite{Dobrescu:2021vak}. The $H_u$ and $H_d$ $SU(2)$ Higgs doublets are adopted to implement the DFSZ model given their couplings to the SM quarks, but instead of one SM gauge singlet scalar to define the Peccei-Quinn symmetry in the scalar potential, we include two SM gauge singlet scalar fields since one Goldstone mode is eaten by the $Z'$ gauge boson.  Hence, the $\Phi_A$ and $\Phi_B$ baryonic Higgs fields provide both the axion degree of freedom in a DFSZ-like manner as well as the longitudinal mode of the $Z'$ boson, in addition to breaking the chiral symmetry of the $U(1)_B$ gauge symmetry and giving the primary source of mass to the anomalon fields.  After assigning the $\mathbb{Z}_4$ discrete charges to the fields and writing down the complete set of allowed Yukawa terms, we enjoy the accidental PQ symmetry displayed in~\tableref{U1B_DFSZ}.
\begin{table}[ht]
\centering
\resizebox{0.5\textwidth}{!}{
\begin{tabular}{c||ccc|c|c||c}
 & $SU(3)_C$ & $SU(2)_L$ & $U(1)_Y$ & $U(1)_B$ & $\mathbb{Z}_4$ & $U(1)_{PQ}$\\ 
 \hline
$Q_L^i$ & 3 & 2 & 1/6 & 1/3 & +1 & $X_Q$ \\
$u_R^i$ & 3 & 1 & 2/3 & 1/3 & -i & $X_Q$-$X_u$ \\
$d_R^i$ & 3 & 1 & -1/3 & 1/3 & +1 & $X_Q$-$X_d$ \\
$L_L^i$ & 1 & 2 & -1/2 & 0 & +1 & $X_L$ \\
$e_R^i$ & 1 & 1 & -1 & 0 & +1 & $X_L$-$X_d$ \\ 
\hline
$H_u$ & 1 & 2 & -1/2 & 0 & +i & $X_u$ \\
$H_d$ & 1 & 2 & 1/2 & 0 & +1 & $X_d$ \\ 
\hline
$L_L'$ & 1 & 2 & -1/2 & -1 & +1 & $X'$ \\
$L_R'$ & 1 & 2 & -1/2 & 2 & +i & $X'$-$X_B$ \\
$E_L'$ & 1 & 1 & -1 & 2 & +i & $X'$-$X_d$-$X_B$ \\
$E_R'$ & 1 & 1 & -1 & -1 & +1 & $X'$-$X_d$ \\
$N_L'$ & 1 & 1 & 0 & 2 & +1 & $X'$-$X_u$-$X_B$ \\
$N_R'$ & 1 & 1 & 0 & -1 & -i & $X'$-$X_u$ \\ 
\hline
$\Phi_A$ & 1 & 1 & 0 & -3 & -1 & $-X_A$ \\
$\Phi_B$ & 1 & 1 & 0 & 3 & +i & $-X_B$ \\
\end{tabular}
}
\caption[Field content for a DFSZ-like model with gauged baryon number]{Field content for a DFSZ-like model with gauged baryon number. In addition to the SM fermions we have two Higgs doublets $H_u$ and $H_d$ for EWSB, two Higgs fields $\Phi_A$ and $\Phi_B$ for breaking baryon number and heavy anomalon fields which cancel gauge anomalies. Besides the SM charges, we assign baryon number and a $\mathbb{Z}_4$ charge. The PQ charge results as global symmetry of the Lagrangian.}
\label{table:U1B_DFSZ}
\end{table}

Our scalar Lagrangian is given by
\begin{align}
\mathcal{L}_\text{scalar} &\supset 
|D_\mu H_u|^2 + |D_\mu H_d|^2 + |D_\mu \Phi_A|^2 + |D_\mu \Phi_B|^2 \nonumber \\
    & - V \left( |H_u|^2 , |H_d|^2 , |\Phi_A|^2 , |\Phi_B|^2 \right) - \lambda_{AB} \left( H_u^{aT}\epsilon^{ab} H_d^b \Phi_A \Phi_B + \text{h.c.} \right) \ ,
\label{eqn:scalar_tot}
\end{align}
where $\lambda_{AB}$ uniquely determines the accidental PQ symmetry.  
We remark that other choices of the discrete symmetry can permit more terms in the scalar potential that explicitly break this prescribed PQ symmetry, but this will be discussed in a separate publication~\cite{Elahi:TBA}.  The accidental PQ charges from~\tableref{U1B_DFSZ} are also realized in the corresponding Yukawa Lagrangian, which has the form
\begin{align}
    \mathcal{L}_\text{Yukawa} \supset
    & - y_u^{ij} \bar{Q}^i_L H_u u^j_R - y_d^{ij} \bar{Q}^i_L H_d d^j_R - y_e^{ij} \bar{L}^i_L H_d e^j_R \nonumber \\
    & - y_L \bar{L}'_R \Phi_B L'_L - y_E \bar{E}'_L \Phi_B E'_R - y_N \bar{N}'_L \Phi_B N'_R \nonumber \\
    & - y_1 \bar{L}'_L H_d E'_R - y_2 \bar{L}'_R H_d E'_L - y_3  \bar{L}'_L H_u N'_R - y_4 \bar{L}'_R H_u N'_L + \text{h.c.} \ .
\label{eqn:yukawa_tot}
\end{align}
Here, the first line reflects the type-II two Higgs doublet Yukawa interactions of the DFSZ model, while the remaining terms are the standard anomalon Yukawa terms arising in gauged baryon number models~\cite{Dobrescu:2013cmh, Dobrescu:2014fca, Michaels:2020fzj, Dobrescu:2021vak}.  We assume all of the scalar fields will acquire vacuum expectation values (vevs) from the $V(|H_u|^2, |H_d|^2, |\Phi_A|^2, |\Phi_B|^2)$ scalar potential, and we will focus on the Goldstone bosons from the corresponding complex scalar fields.

In order to analyze the axion and $Z'$ phenomenology, we now make use of the scalar Lagrangian in~\eqnref{scalar_tot} to define the Goldstone basis of the angular fields and their interactions to the radial fields. The Yukawa Lagrangian in~\eqnref{yukawa_tot} gives us the mass basis of the anomalons and their couplings to scalars.  We will also present the scalar and fermion couplings to gauge bosons, where we include an effective one-loop induced kinetic mixing of the form
\begin{equation}
\mathcal{L}_\text{kin-mix} \supset \frac{\epsilon_\text{eff}}{2} B_{\mu\nu} K^{\mu\nu} ,
\label{eqn:kinmix}
\end{equation}
leading to a shift in the neutral current gauge fields, and $K_\mu$ denotes the additional gauge boson before diagonalization and canonical normalization.

\subsection{Scalar sector}
\label{subsec:scalars}

To define the Goldstone basis of the angular fields, we parameterize the complex fields via
\begin{gather}
    H_u = \frac{1}{\sqrt{2}} e^{i \pi_u^i \sigma^i / v_u} \left( \begin{array}{c} 
        v_u + h_u \\ 0 
    \end{array} \right) \ , \quad 
    H_d = \frac{1}{\sqrt{2}} e^{i \pi_d^i \sigma^i / v_d} \left( \begin{array}{c} 
        0 \\ v_d + h_d
    \end{array} \right) \ , \nonumber\\ 
    \Phi_A = \frac{v_A+h_A}{\sqrt{2}} e^{i a_A / v_A} \ , \quad 
    \Phi_B = \frac{v_B+h_B}{\sqrt{2}} e^{i a_B / v_B} \ . 
\label{eqn:fieldparam}
\end{gather}
Here, $v_i$ denote the vevs of the scalar fields, $h_i$ are the radial modes, and $a_i$ and $\pi_i$ are angular modes, while $\sigma^i$ denote the Pauli matrices as generators of $SU(2)_L$.  We define $a_u \equiv \pi_u^3$, $a_d \equiv -\pi_d^3$ as neutral angular modes of $H_u$ and $H_d$.  The vevs which spontaneously break $SU(2)_L \times U(1)_Y$ and $U(1)_B$ are $v \equiv \sqrt{v_u^2 + v_d^2}$ and $v' \equiv \sqrt{v_A^2 + v_B^2}$, respectively, with $\tan \beta \equiv v_u / v_d$ and $\tan \beta' \equiv v_A / v_B$. 

Since the PQ symmetry is orthogonal to the gauge $U(1)$ symmetries, we have the relations
\begin{equation}
    0 = \sum_{\{H_i \}} Y_i X_i v_i^2 \ , \quad 
    0 = \sum_{\{\Phi_i\}} B_i X_i v_i^2 \ , \quad 
    X^2v_a^2 = \sum_{\{H_i, \Phi_i\}} X_i^2v_i^2 \ .
\label{eqn:orth_rels}
\end{equation}
Together with the requirement of $X\equiv X_u+X_d=X_A+X_B$ to keep the $\lambda_{AB}$ term PQ invariant, the PQ charges of the scalar fields evaluate to
\begin{equation}
    X_u = X \cos^2{\beta} \ , \quad X_d = X\sin^2{\beta} \ , \quad X_A = X \cos^2{\beta'} \ , \quad X_B = X \sin^2{\beta'} \ . 
\end{equation}
Consequently, the PQ symmetry is spontaneously broken by the effective scale $v_a$, where $v_a \equiv \sqrt{v^2 \sin^2 \beta \cos^2 \beta + v'^2 \sin^2 \beta' \cos^2 \beta'}$, with $\tan \gamma \equiv v \sin \beta \cos \beta / (v' \sin \beta' \cos \beta')$. The PQ charge normalization $X$ is then fixed by the axion decay constant $f_a\equiv Xv_a$. 

To identify the axion $a$, the heavy $SU(2)$ pseudoscalar $A_0$, and the two Goldstones $G_0$, $G_B$ for the longitudinal $Z$ and $Z'$ bosons, we perform the following orthogonal transformation,
\begin{align}
    \begin{pmatrix}
        a \\ A_0 \\ G_0 \\ G_B
    \end{pmatrix}=
    \begin{pmatrix}
        c_{\beta} s_\gamma & s_{\beta} s_\gamma & - c_{\beta'} c_\gamma & - s_{\beta'} c_\gamma \\
        c_{\beta} c_\gamma & s_{\beta} c_\gamma & c_{\beta'} s_\gamma & s_{\beta'} s_\gamma \\
        - s_{\beta} & c_{\beta} & 0 & 0 \\
        0 & 0 & - s_{\beta'} & c_{\beta'}
    \end{pmatrix}
    \begin{pmatrix}
        a_u \\ a_d \\ a_A \\ a_B
    \end{pmatrix} \ ,
    \label{eqn:Goldstone_basis}
\end{align}
where the $G_0$ and $G_B$ Goldstones are easily identified as aligning with the Higgs basis of each sector.  We remark that for $v \ll v'$, $\gamma \approx 0$, we reproduce the invisible axion of the DFSZ model which is dominantly composed of $a_A$ and $a_B$ SM gauge singlets.  We can also reproduce the Weinberg-Wilczek model~\cite{Weinberg:1977ma, Wilczek:1977pj} by considering the other limit, $v \gg v'$, $\gamma\approx \pi/2$.  

The heavy pseudoscalar $A_0$ gets a mass from the $\lambda_{AB}$ term given by 
\begin{equation}
    m_{A_0}^2=\frac{\lambda_{AB}}{2} \frac{v_a^2}{s_\beta c_\beta s_{\beta'} c_{\beta'}} \ .
    \label{eqn:mA0}
\end{equation}
A mass for the axion is only induced by instanton effects, which are quantified by the topological susceptibility $\chi$~\cite{GrillidiCortona:2015jxo},
\begin{equation}
    \chi = m_a^2f_a^2 \ , \quad \chi_\text{QCD} = \frac{m_um_d}{(m_u+m_d)^2} m_\pi^2 f_\pi^2 \ .
    \label{eqn:topsus}
\end{equation}
For an ALP, $\chi$ remains a free parameter, while a vanilla QCD axion has $\chi = \chi_{\text{QCD}}$, although recent studies have demonstrated that $\chi$ can be enhanced by non-QCD sources and still preserve the axion solution to the strong CP problem~\cite{Agrawal:2017ksf, Agrawal:2017evu, Gaillard:2018xgk, Kivel:2022emq}.  As long as $\lambda_{AB}$ is sufficiently large, the basis rotation in~\eqnref{Goldstone_basis} coincides with the mass basis of $a$ and $A_0$.

For the $CP$ even Higgs bosons, we perform the orthogonal transformation to the Higgs basis in the alignment limit, giving
\begin{align}
    \begin{pmatrix}
    h\\ H_0\\ h'\\ H_0'
    \end{pmatrix} \equiv
    \begin{pmatrix}
    s_\beta & c_\beta & 0& 0\\
    c_\beta & -s_\beta & 0& 0\\
    0& 0& s_{\beta'} & c_{\beta'}\\
    0& 0& c_{\beta'} & -s_{\beta'}
    \end{pmatrix}
    \begin{pmatrix}
    h_u\\ h_d\\ h_A\\ h_B
    \end{pmatrix} \ .
\end{align}
We will assume that the Higgs basis is aligned with the mass basis and neglect further scalar mixing, since our focus is the phenomenology of the light axion and $Z'$ boson.  Large deviations from the alignment limit are also strongly constrained by Higgs observables~\cite{ATLAS:2022vkf, CMS:2022dwd}.

\subsection{Fermion sector}
\label{subsubsec:fermions}

In this section, we calculate the anomalon masses and couplings to the axion and other scalars.  From~\eqnref{yukawa_tot}, the masses of the anomalons arise from the vevs $v_B$, $v_u$ and $v_d$, where we parameterize the Yukawa couplings by $y_k = |y_k| \exp i \delta_k$.  After accounting for rephasing freedom, we have two complex phases signifying $CP$ violation which we shift into the Yukawa terms with $H_u$ and $H_d$,
\begin{align}
    \mathcal{L}_\text{anom} \supset 
    &-|y_L| \bar{L}'_R \Phi_B L'_L-|y_E| \bar{E}'_L \Phi_B E'_R - |y_1| e^{i\delta_{12}} \bar{L}'_L H_d E'_R-|y_2| e^{-i\delta_{12}} \bar{L}'_R H_d E'_L \nonumber \\
    &-|y_N|\bar{N}'_L\Phi_BN'_R-|y_3|e^{i\delta_{34}} \bar{L}'_L H_u N'_R-|y_4| e^{-i\delta_{34}} \bar{L}'_R H_u N'_L+\text{h.c.} \ ,
\label{eqn:anom_yuk}
\end{align}
where the physical complex phases $\delta_{12}$ and $\delta_{34}$ are given by
\begin{equation}
\delta_{12} = \frac{\delta_1 - \delta_2 + \delta_L - \delta_E}{2} \ , \quad 
\delta_{34} = \frac{\delta_3 - \delta_4 + \delta_L - \delta_N}{2} \ .
\end{equation}
The induced mass parameters are 
\begin{gather}
    m_L = \frac{|y_L|}{\sqrt{2}} c_{\beta'} v' \ , \quad 
    m_E = \frac{|y_E|}{\sqrt{2}} c_{\beta'} v' \ , \quad 
    m_N = \frac{|y_N|}{\sqrt{2}} c_{\beta'} v' \ ,\nonumber\\
    m_1 = \frac{|y_1|}{\sqrt{2}} c_{12} c_{\beta} v \ , \quad 
    m_2 = \frac{|y_2|}{\sqrt{2}} c_{12} c_{\beta} v \ , \quad 
    m_3 = \frac{|y_3|}{\sqrt{2}} c_{34} s_{\beta} v \ , \quad 
    m_4 = \frac{|y_4|}{\sqrt{2}} c_{34} s_{\beta} v \ ,
\end{gather}
where $c_{12} \equiv \cos \delta_{12}$ and $c_{34} \equiv \cos \delta_{34}$ reflect the impact of the CP violating phases.
We introduce the shorthand
\begin{equation}
m_{ij} \equiv \frac{m_i+m_j}{2} \ , \quad 
\Delta_{ij} \equiv \frac{m_i-m_j}{m_i+m_j} \ , \quad 
t_{ij} \equiv \tan{\delta_{ij}} \ ,
\label{eqn:m_def}
\end{equation}
after which the mass mixing matrix becomes
\begin{align}
    \mathcal{L}_\text{anom} \supset 
    &-\begin{pmatrix}
    \bar{e}'_L & \bar{E}'_L
    \end{pmatrix}
    \begin{pmatrix}
    m_L & m_{12}(1+\Delta_{12})(1+it_{12})\\m_{12}(1-\Delta_{12})(1-it_{12}) & m_E
    \end{pmatrix}
    \begin{pmatrix}
    e'_R\\E'_R
    \end{pmatrix}
    \label{eqn:fermionmassgeneral} \\
    &-\begin{pmatrix}
    \bar{\nu}'_L & \bar{N}'_L
    \end{pmatrix}
    \begin{pmatrix}
    m_L & m_{34}(1+\Delta_{34})(1+it_{34})\\m_{34}(1-\Delta_{34})(1-it_{34}) & m_N
    \end{pmatrix}
    \begin{pmatrix}
    \nu '_R\\N'_R
    \end{pmatrix}
    +\text{h.c.} \ . \nonumber
\end{align}
For $v'\gg v$, the off-diagonal terms are at least suppressed by $v/v_a$. The $\Delta_{12}$ and $\Delta_{34}$ terms are also suppressed by the difference in the Yukawa couplings, which is generally negligible unless the couplings are hierarchical, and so we will assume $\Delta_{12} = \Delta_{34} = 0$ for the remainder of this work.

The $CP$ violation is encoded via the tangent of the $CP$ violating phases and will cause mixing between the axion $a$ with the SM Higgs $h$.  Since we are aligned in the Higgs basis, we will set $t_{12} = t_{34} = 0$ and leave a study of small deviations inducing mixing between $a$ and $h$ to future work.

After these simplifying assumptions, we can now rotate the symmetric mass matrices of the anomalons in~\eqnref{fermionmassgeneral} using $\alpha_E$ and $\alpha_N$ mixing angles defined via 
\begin{align}
    \begin{pmatrix}
        E_1\\E_2\end{pmatrix} = 
    \begin{pmatrix} 
        \cos{\alpha_E} & \sin{\alpha_E}\\
        -\sin{\alpha_E} & \cos{\alpha_E}
    \end{pmatrix}
    \begin{pmatrix}
        e' \\ E'
    \end{pmatrix} \ , \quad
    \begin{pmatrix}
        N_1\\N_2
    \end{pmatrix} = 
    \begin{pmatrix}
        \cos{\alpha_N}&\sin{\alpha_N}\\
        -\sin{\alpha_N}&\cos{\alpha_N}
    \end{pmatrix}
    \begin{pmatrix}
        \nu '\\N'
    \end{pmatrix} \ .
\label{eqn:anom_flav}
\end{align}
The masses of $E_1$, $E_2$, $N_1$ and $N_2$ are given by \begin{equation}
    m_{E_{1,2}} = m_{LE} \biggl( 1 \mp \sqrt{\Delta_{LE}^2 + \frac{m_{12}^2}{m_{LE}^2}} \biggr) \ , \quad
    m_{N_{1,2}} = m_{LN} \biggl( 1 \mp \sqrt{\Delta_{LN}^2 + \frac{m_{34}^2}{m_{LN}^2}} \biggr) \ , 
\label{eqn:anom_masses}
\end{equation}
using the shorthand in~\eqnref{m_def}. 

We now evaluate the couplings of the axion field $a$ and the SM Higgs field $h$ to the anomalons. The couplings to the axion are at order $1/f_a$
\begin{align}
    \mathcal{L}_{\text{anom}, \ a} &\supset
        i X_B \frac{a}{f_a} \cos(2 \alpha_E) (m_{E_1} \bar{E}_1 \gamma_5 E_1 - m_{E_2} \bar{E}_2 \gamma_5 E_2 ) \nonumber \\
        & + i X_B \frac{a}{f_a} \sin(2 \alpha_E) \frac{m_{E_1} + m_{E_2}}{2} (\bar{E}_1 \gamma_5 E_2 + \bar{E}_2 \gamma_5 E_1) \nonumber \\
        & + i X_B \frac{a}{f_a} \cos(2\alpha_N) (m_{N_1} \bar{N}_1 \gamma_5 N_1 - m_{N_2} \bar{N}_2 \gamma_5 N_2) \nonumber\\
        & + i X_B \frac{a}{f_a} \sin(2 \alpha_N) \frac{m_{N_1} + m_{N_2}}{2} (\bar{N}_1 \gamma_5 N_2 + \bar{N}_2 \gamma_5 N_1 ) \nonumber \\
        & + i X_d \frac{a}{f_a}  \sin(2 \alpha_E) \frac{m_{E_1} - m_{E_2}}{2} ( \bar{E}_1 E_2 - \bar{E}_2 E_1) \nonumber \\
        & + i X_u \frac{a}{f_a} \sin(2 \alpha_N) \frac{m_{N_1} - m_{N_2}}{2} (\bar{N}_1 N_2 - \bar{N}_2 N_1) \ .
\label{eqn:ax_coup}
\end{align}
Here, we see that the terms proportional to $X_B$ are the canonical axial couplings proportional to fermion masses, while the remaining terms proportional to $X_d$ or $X_u$ scale as the difference of fermion masses and arise generically in flavor violating axion models, as we will discuss in Subsection~\ref{subsec:flavor_changing}.

Separately, the interactions of the SM Higgs $h$ to the anomalons 
are
\begin{align}
    \mathcal{L}_{\text{anom}, \ h} & \supset 
        \sin(2 \alpha_E) \frac{m_{E_1} - m_{E_2}}{2} \frac{h}{v} (\cos(2 \alpha_E) (\bar{E}_1 E_2 + \bar{E}_2 E_1) - \sin(2 \alpha_E) (\bar{E}_1 E_1 - \bar{E}_2 E_2 )) \nonumber \\
        & + \sin(2 \alpha_N) \frac{m_{N_1} - m_{N_2}}{2} \frac{h}{v} (\cos(2 \alpha_N) (\bar{N}_1 N_2 + \bar{N}_2 N_1 ) - \sin(2 \alpha_N) (\bar{N}_1 N_1 - \bar{N}_2 N_2)) \ .
\label{eqn:h_coup}
\end{align}
At dimension 5 we also get a mixed operator 
\begin{align}
        \mathcal{L}_{\text{anom},\ a, \ h}&\supset i X_d \sin(2 \alpha_E) \frac{m_{E_1} - m_{E_2}}{2} \frac{h}{v} \frac{a}{f_a} (\bar{E}_1 E_2 - \bar{E}_2 E_1 )\nonumber\\
        &+ i X_u \sin(2 \alpha_N) \frac{m_{N_1} - m_{N_2}}{2} \frac{h}{v} \frac{a}{f_a} (\bar{N}_1 N_2 - \bar{N}_2 N_1 ) \ .
\label{eqn:a_h_coup}
\end{align}
In the case with $CP$ violation, the linear Higgs interactions would mix with the linear axion interactions proportional to $X_u$ and $X_d$. The last two terms are due to the fact that the interactions of the axion proportional to $X_u$ and $X_d$ are induced by the Higgs doublets and are needed for a complete set of operators at order $1/f_a$. 

\subsection{Gauge sector}
\label{subsubsection:gauge}

Finally, we discuss the $Z$ and $Z'$ interactions, which necessarily includes kinetic mixing effects from~\eqnref{kinmix}. The effective kinetic mixing parameter $\epsilon_\text{eff}$ is determined by calculating the one-loop contribution to the two point interaction between the hypercharge gauge field $B_\mu$ and baryon number gauge field $K_\mu$, giving 
\begin{equation}
\mathcal{L} \supset \frac{\epsilon_\text{eff}}{2} B_{\mu\nu} K^{\mu\nu} + \frac{m_\text{eff}^2}{2} B_{\mu} K^\mu \ ,
\end{equation}
where $m_\text{eff}$ corresponds to a possible mass mixing.  The mass mixing vanishes if all fermions in the loop are vector-like under one of the $U(1)$ gauge symmetries~\cite{Dobrescu:2021vak}.  The divergence in the two-point loop diagram is cancelled after imposing the trace condition on the mediator fermions,
\begin{equation}
    \sum_f N_f(Y_{V}^f B_{V}^f + Y_{A}^f B_{A}^f) = 0 \ ,
\label{eqn:trace_cond}
\end{equation}
where $N_f$ denotes the multiplicity factor of fermion $f$. In the unbroken phase of electroweak symmetry, we can consider the SM fermions to be massless, such that the only remaining contribution comes from the anomalons. For $m_L^2\ ,\ m_E^2\gg p^2$ the effective kinetic mixing parameter reads
\begin{align}
\epsilon_\text{eff} &= \frac{g_Y g_B}{(4\pi)^2} \frac{4}{3} \sum_{f \in \{L',E',N'\}}( Y_{V}^f B_{V}^f + Y_{A}^f B_{A}^f )\biggl( \frac{5}{3} + \ln(\frac{m_f^2}{p^2}) + \mathcal{O}\left(\frac{p^2}{m_f^2}\right)\biggr) \nonumber \\
    & = - \frac{eg_B}{c_W} \frac{1}{(4\pi)^2} \frac{2}{3} \biggl(\frac{10}{3} + \ln(\frac{m_L^2}{p^2}) + \ln(\frac{m_E^2}{p^2}) + \mathcal{O} \left( \frac{p^2}{m_L^2 , m_E^2} \right) \biggr) \ .
\label{eqn:eps_YB}
\end{align}
The large logarithm in $\epsilon_\text{eff}$ cancels roughly the loop factor such that the dominant parametric dependence is given by $\epsilon_\text{eff}\approx eg_Bc_W^{-1}$. In the following we will see that we get new interactions proportional to $\epsilon_\text{eff}$.

We recall from Ref.~\cite{Liu:2017lpo} that kinetic mixing is removed by shifting the gauge fields into a diagonal and canonically normalized basis, using the replacement rule 
\begin{align}
    Z^\text{SM}_\mu = & Z_\mu - \epsilon_\text{eff} s_W \frac{m_{Z'}^2}{m_{Z'}^2 - m_Z^2} Z'_\mu + \mathcal{O} \left( \epsilon_\text{eff}^2 \right)  \ ,
    \label{eqn:shift1}\\
    K_\mu = & Z'_\mu - \epsilon_\text{eff} s_W \frac{m_{Z}^2}{m_{Z}^2 - m_{Z'}^2} Z_\mu + \mathcal{O} \left( \epsilon_\text{eff}^2 \right) \ ,
    \label{eqn:shift2}
\end{align}
to shift to the mass basis.  Assuming $m_{K} > m_{Z, \text{SM}}$, the corresponding masses are
\begin{align}
m_Z &= m_{Z,\text{SM}} \left( 1 + \frac{\epsilon_\text{eff}^2}{2} \frac{s_W^2 m_{Z,\text{SM}}^2 }{ m_{Z,\text{SM}}^2 - m_{K}^2} + \mathcal{O}(\epsilon_\text{eff}^4) \right) \\
m_{Z'} &= m_{K} \left( 1 + \frac{\epsilon_\text{eff}^2}{2} \frac{(m_{K}^2 - c_W^2 m_{Z,\text{SM}}^2)} {m_{K}^2 - m_{Z,\text{SM}}^2 } + \mathcal{O} (\epsilon_\text{eff}^4) \right) \ ,
\end{align}
with $s_W$, $c_W$ being sine and cosine of the weak angle $\theta_W$. We see that the mass correction only appears at order $\epsilon_\text{eff}^2$ and is hence typically negligible.

We apply the shifts in the gauge bosons in~\eqnref{shift1} and~\eqnref{shift2} and obtain for the scalar Lagrangian
\begin{align}
    \mathcal{L}_\text{scalar}^{d\leq 4} \supset 
        & \frac{1}{2} \partial_\mu h \partial^\mu h + \frac{1}{2} \partial_\mu H_0 \partial^\mu H_0 + \frac{1}{2} \partial_\mu h' \partial^\mu h' + \frac{1}{2} \partial_\mu H'_0 \partial^\mu H'_0 - V(h , H_0 , h' , H'_0) \nonumber \\
        & + \frac{1}{2} \partial_\mu a \partial^\mu a + \frac{1}{2} \partial_\mu A_0 \partial^\mu A_0 - \frac{m_{A_0}^2}{2} A_0^2 + \frac{1}{ s_\gamma^2 c_\gamma^2} \frac{m_{A_0}^2}{v_a^2} \frac{A_0^4}{4!} \nonumber \\
        & + \frac{1}{8} \frac{e^2}{s_W^2c_W^2} \left( (h+v)^2+H_0^2 \right) \left( Z_\mu - \epsilon_\text{eff} s_W \frac{m_{Z'}^2}{m_{Z'}^2-m_Z^2} Z'_\mu \right) \left( Z^\mu - \epsilon_\text{eff} s_W \frac{m_{Z'}^2}{m_{Z'}^2 - m_Z^2} Z'^\mu \right) \nonumber\\
        & + \frac{9}{2} g_B^2 \left( (h'+v')^2 + H_0'^2 \right) \left( Z'_\mu - \epsilon_\text{eff} s_W \frac{m_{Z}^2}{m_{Z}^2 - m_{Z'}^2} Z_\mu \right) \left( Z'^\mu - \epsilon_\text{eff} s_W \frac{m_{Z}^2}{m_{Z}^2 - m_{Z'}^2}Z^\mu \right) \nonumber \\
        & - \frac{e}{s_W c_W} H_0 \left( Z_\mu - \epsilon_\text{eff} s_W \frac{m_{Z'}^2}{m_{Z'}^2 - m_Z^2} Z'_\mu \right)( s_\gamma \partial^\mu a + c_\gamma \partial^\mu A_0 ) \nonumber \\
        & + 6 g_B H_0' \left( Z'_\mu - \epsilon_\text{eff} s_W \frac{m_{Z}^2}{m_{Z}^2 - m_{Z'}^2} Z_\mu \right)( c_\gamma \partial^\mu a - s_\gamma \partial^\mu  A_0) + \mathcal{O} \left(\epsilon_\text{eff}^2\right) \ ,
\label{eqn:aZh}
\end{align}
where $G_0$ is absorbed by $Z^\text{SM}_\mu$, $G_B$ by $K_\mu$, and $A_0$ is the only angular mode which gets a mass from the term proportional to $\lambda_{AB}$ as defined in~\eqnref{mA0}.

Finally, we discuss the gauge interactions of the anomalons.  Following Ref.~\cite{Liu:2017lpo}, the current interactions of the neutral gauge bosons are given at $\mathcal{O}( \epsilon_{\text{eff}})$ by\footnote{In contrast to Refs.~\cite{Dobrescu:2013cmh, Dobrescu:2021vak}, our convention for $g_B$ in this work uses $\mathcal{L} = \frac{1}{3} g_B Z'_{\mu} \left( \bar{q} \gamma^\mu q \right)$, and thus our $g_B$ is half the value used in Refs.~\cite{Dobrescu:2013cmh, Dobrescu:2021vak}.}.
\begin{align}
\mathcal{L}_\text{gauge} &\supset eA_\mu J^\mu_Q+\frac{e}{\sqrt{2} s_W}(W_\mu^-J^{+\mu}_W+\text{h.c.})+Z_\mu \left(\frac{e}{s_W c_W} J^\mu_Z-\epsilon_\text{eff} s_W g_B \frac{m_Z^2}{m_Z^2-m_{Z'}^2}J^\mu_B\right)\nonumber\\
& +Z'_\mu \left(g_B J^\mu_B+\epsilon_\text{eff}eJ^\mu_Q-\epsilon_\text{eff} \frac{e}{c_W} \frac{m_{Z'}^2}{m_{Z'}^2-m_Z^2}J^\mu_Z\right) \ , 
\end{align}
with the gauge currents given by
\begin{align}
    J^\mu_Q \supset & - (\bar{E}_1 \gamma^\mu E_1 + \bar{E}_2 \gamma^\mu E_2) \ , \\
    J^{+\mu}_W \supset & (c_E c_N \bar{E}_1 \gamma^\mu  N_1 + c_E s_N \bar{E}_1 \gamma^\mu N_2 + s_E c_N \bar{E}_2 \gamma^\mu N_1 + s_E s_N \bar{E}_2 \gamma^\mu N_2) \ , \\
    J^\mu_Z \supset & \frac{1}{2}((2 s_W^2 - c_E^2) \bar{E}_1 \gamma^\mu E_1 + (2 s_W^2 - s_E^2) \bar{E}_2 \gamma^\mu E_2 - s_E c_E (\bar{E}_1 \gamma^\mu E_2 + \bar{E}_2 \gamma^\mu E_1)) \nonumber \\
    & + \frac{1}{2} (c_N^2 \bar{N}_1 \gamma^\mu N_1 + s_N^2 \bar{N}_2 \gamma^\mu N_2 + s_N c_N (\bar{N}_1 \gamma^\mu N_2 + \bar{N}_2 \gamma^\mu N_1)) \ , \\
    J^\mu_B \supset & \frac{1}{2} (\bar{E}_1 \gamma^\mu E_1 + \bar{E}_2 \gamma^\mu E_2 + \bar{N}_1 \gamma^\mu N_1 + \bar{N}_2 \gamma^\mu  N_2) \nonumber \\
    & + \frac{3}{2} (\cos(2 \alpha_E)(\bar{E}_1 \gamma^\mu \gamma_5 E_1 - \bar{E}_2 \gamma^\mu \gamma_5 E_2) + \sin(2 \alpha_E) (\bar{E}_1 \gamma^\mu \gamma_5 E_2 + \bar{E}_2 \gamma^\mu \gamma_5 E_1)) \nonumber \\
    & + \frac{3}{2}(\cos(2 \alpha_N) (\bar{N}_1 \gamma^\mu \gamma_5 N_1 - \bar{N}_2 \gamma^\mu \gamma_5 N_2) + \sin(2 \alpha_N) (\bar{N}_1 \gamma^\mu \gamma_5 N_2 + \bar{N}_2 \gamma^\mu \gamma_5 N_1)) \ .
\end{align}
There are two limiting cases: $\alpha_i\to 0$ corresponds to minimal mixing, while $\alpha_i\to\pi /4$ describes maximal mixing. In the minimal mixing case, we recover flavor-conserving axion and $Z'$ couplings, while in the maximal mixing case, the axion, $Z$ and $Z'$ bosons all change the flavor of the anomalons.  This will be further discussed in~\subsecref{flavor_changing}.  Another feature is given by the fact that the anomalons give rise to new contributions to the Higgs decay to two gauge bosons which are not excluded~\cite{Michaels:2020fzj}. 

\section{Flavor Dependent Basis Transformations of Axion and $Z'$ EFT Operators}
\label{sec:EFT}

In this section, we construct the low energy effective field theory of the axion and $Z'$ boson at energy scales well below the anomalon masses.  We are particularly interested in the general structure of flavor-conserving and flavor-violating axion interactions and how they manifest when the mediator fermions are integrated out.  For this purpose, we consider a chiral transformation of the fermion fields which accounts for possible flavor-violating effects.  The chiral transformation is also relevant for understanding the basis invariance of flavor-conserving and flavor-violating axion interactions.  
We represent the general axion interaction basis in a form where the axion appears in the Yukawa and gauge interactions of the fermions.  We include a dimension 5 commutator interaction for the axion coupling to fermions and a gauge boson, which only appears when the PQ and the gauge currents are flavor-violating.  This commutator plays a crucial role in maintaining the basis invariance of the effective coupling between the axion, Higgs boson, and a gauge boson.  Apart from the commutator interaction, our basis is equivalent to the commonly used operator set from Ref.~\cite{Georgi:1986df}.

\subsection{General axion interaction basis}
\label{subsec:flavor_changing}

Our primary goal in this subsection is constructing a complete basis set of operators for axion interactions appropriate for characterizing flavor conserving and flavor violating axion interactions.  We begin by writing the fermions in~\eqnref{anom_flav} as a vector $\psi$ with a diagonal mass matrix $\mathbf{M}^\psi$.  Correspondingly, from~\eqnref{ax_coup}, we identify the terms proportional to $X_B$ with the anti-commutator $\{\mathbf{M}^\psi,\mathbf{X}_A^\psi \}$, where $\mathbf{X}_A^\psi$ is the so-called axial PQ charge matrix, and we identify the terms proportional to $X_d$ and $X_u$ with the commutator $[\mathbf{M}^\psi,\mathbf{X}_V^\psi]$ for the vector-like PQ charge matrix $\mathbf{X}_V^\psi$.  For the $E$ anomalons, the axial and vector-like PQ charge matrices are $\mathbf{X}_{A}^E = X_B \mathbf{C}_E$ and $\mathbf{X}_{V}^E = X_d \mathbf{C}_E$, while the $N$ anomalons have equivalent expressions $\mathbf{X}_{A}^N = X_B \mathbf{C}_N$ and $\mathbf{X}_{V}^N = X_u \mathbf{C}_N$, where 
\begin{equation}
    \mathbf{C}_E = \frac{1}{2} 
        \begin{pmatrix}
            \cos(2\alpha_E) &\sin(2\alpha_E)\\
            \sin(2\alpha_E) &-\cos(2\alpha_E)
        \end{pmatrix}, \quad
    \mathbf{C}_N = \frac{1}{2}
    \begin{pmatrix}
        \cos(2\alpha_N) &\sin(2\alpha_N)\\
        \sin(2\alpha_N) &-\cos(2\alpha_N)
    \end{pmatrix} \ .
\label{eqn:X_rotation}
\end{equation}
We remark that the diagonal entries of $\mathbf{X}_V^\psi$ do not contribute to the commutator with $\mathbf{M}^\psi$ but are chosen such that $\mathbf{X}_V^\psi$ commutes with $\mathbf{X}_A^\psi$. 

Including the interactions with the scalar fields $\phi_K$ in the Higgs basis, we can write the Yukawa interaction Lagrangian at order $1/f_a$ as
\begin{align}
    \mathcal{L}_{\psi,\text{Yuk}} & \supset -\sum_K\left(
         \frac{v_K+\phi_K}{\sqrt{2}} \bar{\psi} \mathbf{Y}_K^\psi \psi - i \frac{a}{f_a} \frac{v_K+\phi_K}{\sqrt{2}}\bar{\psi} \{\mathbf{Y}_K^\psi , \mathbf{X}_A^\psi\} \gamma_5 \psi -  i \frac{a}{f_a} \frac{v_K+\phi_K}{\sqrt{2}}\bar{\psi} [\mathbf{Y}_K^\psi , \mathbf{X}_V^\psi] \psi\right)  \ ,
\label{eqn:LY1}
\end{align}
where the Yukawa matrices $\mathbf{Y}_K^\psi$ are defined in the mass basis of the fermions via $\mathbf{M}^\psi\equiv\sum_K\mathbf{Y}_K^\psi v_K/\sqrt{2}$.  This structure is equivalent to the first-order approximation of an exponential interaction,
\begin{align}
    \mathcal{L}_{\psi,\text{Yuk}} & \simeq  -\sum_K  \frac{v_K+\phi_K}{\sqrt{2}}\bar{\psi} \exp \left( i (\mathbf{X}_V^\psi - \mathbf{X}_A^\psi \gamma_5 ) \frac{a}{f_a} \right) \mathbf{Y}_K^\psi \exp \left( - i (\mathbf{X}_V^\psi + \mathbf{X}_A^\psi \gamma_5) \frac{a}{f_a} \right) \psi \ .
\label{eqn:LY2}
\end{align}
We note that the ordering of $\mathbf{X}_V^\psi$ and $\mathbf{X}_A^\psi$ in the exponent is free since both matrices commute.  

In order to identify all possible axion interactions which originate from the most general fermion Lagrangian, we now include the covariant derivatives of the fermions.  The general Lagrangian describing all terms which involve $\psi$ is then given by
\begin{align}
    \mathcal{L}_\psi &\supset 
         i \bar{\psi} \gamma^\mu \partial_\mu \psi  \nonumber \\
        & - \sum_K\frac{v_K+\phi_K}{\sqrt{2}}\bar{\psi} \exp \left( i (\mathbf{X}_V^\psi - \mathbf{X}_A^\psi \gamma_5) \frac{a}{f_a} \right) \mathbf{Y}_K^\psi \exp \left( - i (\mathbf{X}_V^\psi + \mathbf{X}_A^\psi \gamma_5) \frac{a}{f_a} \right) \psi\nonumber\\
        &+ \sum_I g_I \bar{\psi} A_\mu^I \gamma^\mu (\mathbf{Q}_{IV}^\psi + \mathbf{Q}_{IA}^\psi \gamma_5) \psi \ .
\label{eqn:Lpsi1}
\end{align}
Here, $I$ denotes the gauge group and $A_\mu^I=A_\mu^{Ia}T^a_I$ the corresponding gauge bosons. The gauge interactions of the fermions are defined by a vector-like and an axial charge matrix, $\mathbf{Q}_{IV}^\psi$ and $\mathbf{Q}_{IA}^\psi$, which would include the CKM matrix in the case of $SU(2)_L$ gauge bosons.

We have now specified all of the axion couplings at dimension 4 and dimension 5 defined by our model.  At one-loop, these couplings will induce axion interactions with gauge bosons at the same order in $1 / f_a$.  Moreover, following Fujikawa's derivation~\cite{Fujikawa:1979ay} of the Adler-Bell-Jackiw $U(1)$ chiral anomaly, axial phase rotations of the fermions will also cause shifts in the axion couplings to gauge bosons.

Since~\eqnref{LY1} encodes the entire flavor structure of the axion coupling to fermions, we can perform the chiral transformation 
\begin{equation}
    \psi \to \exp \left( i (\mathbf{X}_V^\psi + \mathbf{X}_A^\psi \gamma_5) \frac{a}{f_a} \right) \psi , \quad
    \bar{\psi} \to \bar{\psi} \exp \left( - i (\mathbf{X}_V^\psi - \mathbf{X}_A^\psi \gamma_5) \frac{a}{f_a} \right)
\label{eqn:trafo_general}
\end{equation}
for each type of fermion to generate a set of operators at order $1/f_a$ which is closed under chiral fermion transformations.  This transformation removes the axion couplings from the Yukawa interactions and shifts them into the gauge interactions and a derivative interaction,
\begin{align}
    \mathcal{L}_\psi &\to 
    i \bar{\psi} \gamma^\mu \partial_\mu \psi - \frac{\partial_\mu a}{f_a} \bar{\psi} \gamma^\mu (\mathbf{X}_V^\psi + \mathbf{X}_A^\psi \gamma_5) \psi\nonumber\\
    &- \bar{\psi} \mathbf{M}^\psi \psi - \sum_K \frac{\phi_K}{\sqrt{2}} \bar{\psi} \mathbf{Y}_K^\psi \psi  \nonumber \\
    & + \sum_I g_I \bar{\psi} A_\mu^I \gamma^\mu \exp \left( - i (\mathbf{X}_V^\psi + \mathbf{X}_A^\psi \gamma_5) \frac{a}{f_a} \right) (\mathbf{Q}_{IV}^\psi + \mathbf{Q}_{IA}^\psi \gamma_5) \exp \left( i (\mathbf{X}_V^\psi + \mathbf{X}_A^\psi \gamma_5) \frac{a}{f_a} \right) \psi \nonumber \\
    &+ \frac{a}{f_a} \sum_{I,J} \mathcal{A}_{\text{PQ} IJ} \frac{g_I g_J}{(4\pi)^2} F^a_{I\mu\nu} \tilde{F}^{a,\mu\nu}_J \ .
\label{eqn:Lpsi2}
\end{align}
The anomalous coupling to gauge bosons in the last row is determined by Fujikawa's method \cite{Fujikawa:1979ay}. It is proportional to the anomaly coefficient $\mathcal{A}_{\text{PQ}IJ}$ given by
\begin{align}
    \mathcal{A}_{\text{PQ}IJ} 
        & = \sum_{i,j,k} T(R_{Ik}) (X_R^{ij} Q_{IR}^{ik} Q_{JR}^{kj} - X_L^{ij} Q_{IL}^{ik} Q_{JL}^{kj} ) \nonumber \\
        & = 2 \sum_{i,j,k} T(R_{Ik}) (X_V^{ij} (Q_{IV}^{ik} Q_{JA}^{kj} + Q_{IA}^{ik} Q_{JV}^{kj}) + X_A^{ij} (Q_{IV}^{ik} Q_{JV}^{kj} + Q_{IA}^{ik} Q_{JA}^{kj})) \ ,
\end{align}
where $T(R_{If})$ is the Dynkin index given by $\tr[T_R^a T_R^b] \equiv T(R) \delta^{ab}$~\cite{DiLuzio:2020wdo} with $R_{If}$ denoting the representation of fermion $f$ under gauge group $I$.  In case of a $U(1)$ gauge symmetry, it simply counts the multiplicity.

We now focus on the third line of~\eqnref{Lpsi2} with exponential factors in the gauge boson couplings to axions.  At leading order in $1 / f_a$, these interactions are 
\begin{align}
    \mathcal{L}_{\psi,\text{gauge}} \to 
        &  \sum_I g_I \bar{\psi} A_\mu^I \gamma^\mu (\mathbf{Q}_{IV}^\psi + \mathbf{Q}_{IA}^\psi \gamma_5) \psi + i \frac{a}{f_a} \sum_I g_I \bar{\psi} A_\mu^I \gamma^\mu [\mathbf{Q}_{IV}^\psi + \mathbf{Q}_{IA}^\psi \gamma_5 , \mathbf{X}_V^\psi + \mathbf{X}_A^\psi \gamma_5 ] \psi  \ .
\label{eqn:comm_int}
\end{align}
We see that the axial transformation from~\eqnref{trafo_general} induces a current which contains the commutator of the gauge charge matrix and the PQ charge matrix, defined by
\begin{align}
     J^{\mu,a}_{[ I , \text{PQ}] } & \equiv \sum_\psi\bar{\psi} T^a_I \gamma^\mu [\mathbf{Q}_{IV}^\psi + \mathbf{Q}_{IA}^\psi \gamma_5 , \mathbf{X}_V^\psi + \mathbf{X}_A^\psi \gamma_5] \psi \nonumber \\
    & = \sum_{i,j,k} \bar{\psi}_i T^a_I \gamma^\mu (Q_{IV}^{ik} + Q_{IA}^{ik} \gamma_5 ) ( X^{kj}_V + X^{kj}_A \gamma_5) \psi_j + \text{h.c.} \ .
\end{align}
We note the current is only non-vanishing if both charge matrices are not flavor-conserving. 

Using our results from this section, we now have the general axion interaction basis at order $1/f_a$ including possible flavor-violation effects,
\begin{align}
    \mathcal{L}_{\psi,\text{axion}} \supset 
        &  -\frac{\partial_\mu a}{2f_a} \bar{\psi} \gamma^\mu (\mathbf{C}_{1V}^\psi + \mathbf{C}_{1A}^\psi \gamma_5) \psi + \frac{i}{2} \frac{a}{f_a} \bar{\psi} [\mathbf{M}^\psi , \mathbf{C}_{2V}^\psi] \psi + \frac{i}{2} \frac{a}{f_a} \bar{\psi} \{\mathbf{M}^\psi , \mathbf{C}_{2A}^\psi\} \gamma_5 \psi \nonumber \\
        & + \frac{i}{2} \frac{a}{f_a} \sum_K \frac{\phi_K}{\sqrt{2}} \bar{\psi} [\mathbf{Y}_K^\psi , \mathbf{C}_{KV}^\psi] \psi + \frac{i}{2} \frac{a}{f_a} \sum_K \frac{\phi_K}{\sqrt{2}} \bar{\psi} \{\mathbf{Y}_K^\psi , \mathbf{C}_{KA}^\psi\}\gamma_5 \psi \nonumber \\
        & - \frac{i}{2} \frac{a}{f_a} \sum_I g_I \bar{\psi} A_\mu^I \gamma^\mu [\mathbf{Q}_{IV}^\psi + \mathbf{Q}_{IA}^\psi \gamma_5 , \mathbf{C}_{IV}^\psi + \mathbf{C}_{IA}^\psi \gamma_5] \psi  + \frac{a}{f_a} \sum_{I,J} C_3^{IJ} \frac{g_I g_J}{(4\pi)^2} F^a_{I\mu\nu} \tilde{F}^{a,\mu\nu}_J \ .
\label{eqn:ALP_coup_flavor}
\end{align}
This Lagrangian now describes a closed set of operators under chiral fermion transformations. We remark the first term of the third line is a new dimension-5 operator which couples the axion and a gauge boson to a fermion current that only appears for flavor-changing fermion interactions.  We will see in Subsection~\ref{subsec:avs} that all of these terms are necessary as a basis invariant description of the effective axion coupling to a gauge boson and a scalar boson. In particular, the closure of the operator basis in~\eqnref{ALP_coup_flavor} can be seen by noting that a further general axial transformation from~\eqnref{trafo_general} shifts the Lagrangian couplings in~\eqnref{ALP_coup_flavor} via
\begin{align}
    \mathbf{C}_{1V/A}^\psi&\to\mathbf{C}_{1V/A}^\psi+2\mathbf{X}_{V/A}^\psi\ ,\ \nonumber
    \mathbf{C}_{2V/A}^\psi\to\mathbf{C}_{2V/A}^\psi-2\mathbf{X}_{V/A}^\psi\ ,\  \\
    \mathbf{C}_{KV/A}^\psi &\to\mathbf{C}_{KV/A}^\psi-2\mathbf{X}_{V/A}^\psi\ ,\  
    \mathbf{C}_{IV/A}^\psi\to\mathbf{C}_{IV/A}^\psi-2\mathbf{X}_{V/A}^\psi\ , \nonumber \\
    C_3^{IJ} &\to C_3^{IJ}+\mathcal{A}_{\text{PQ}IJ}  \ . \label{eqn:trafo_changes} 
\end{align}
Having established the framework for axion interactions closed under axial transformations, we now apply our result to the fermions, vectors and scalars from our model.

\subsection{Axion and $Z'$ EFT Lagrangian}
\label{subsec:axionEFTLag}

Having established the required complete set of operators, we now construct the explicit axion and $Z'$ EFT Lagrangian for our model.  Our EFT is generated by integrating out the chiral $U(1)_B$ anomalon content, while the SM fields, axion, and $Z'$ boson remain dynamical.

For the SM fermions, we apply the basis transformation from~\eqnref{trafo_general} such that the axion interacts via the canonical derivative coupling
\begin{equation}
    \mathcal{L}_\text{deriv}=- \frac{\partial_\mu a}{f_a}\sum_{\psi\in\text{SM}} \bar{\psi} \gamma^\mu (\mathbf{X}_V^\psi + \mathbf{X}_A^\psi \gamma_5) \psi\equiv \frac{\partial_\mu a}{ f_a} J^\mu_\text{PQ, SM}\ .
\end{equation}
Separately, integrating out the anomalons generates the following effective axion and $Z'$ Lagrangian at order $1/f_a$,
\begin{align}
    \mathcal{L}_\text{axion} \supset & 
        + \frac{1}{2} (\partial_\mu a) (\partial^\mu a) - \frac{m_a^2}{2} a^2 + \frac{\partial_\mu a}{ f_a} J^\mu_\text{PQ,SM}- C^\text{eff}_{Zh} h Z_\mu \partial^\mu a - C^\text{eff}_{Z'h} h Z'_\mu \partial^\mu a\nonumber\\
        &+ \left(C^\text{SM}_{\gamma \gamma}+C^\text{eff}_{\gamma \gamma}\right) \frac{e^2}{(4\pi)^2} \frac{a}{f_a} F_{\mu\nu} \tilde{F}^{\mu\nu}  + \left(C^\text{SM}_{Z \gamma}+C^\text{eff}_{Z \gamma}\right) \frac{e^2}{s_W c_W} \frac{1}{(4\pi)^2} \frac{a}{f_a} Z_{\mu\nu} \tilde{F}^{\mu\nu}\nonumber \\
        & + \left(C^\text{SM}_{ZZ}+C^\text{eff}_{ZZ}\right) \frac{e^2}{s_W^2 c_W^2} \frac{1}{(4\pi)^2} \frac{a}{f_a} Z_{\mu\nu} \tilde{Z}^{\mu\nu}+ \left(C^\text{SM}_{Z' \gamma}+C^\text{eff}_{Z' \gamma}\right) \frac{g_Be}{(4\pi)^2} \frac{a}{f_a} Z'_{\mu\nu} \tilde{F}^{\mu\nu} \nonumber \\
        & + \left(C^\text{SM}_{Z'Z'}+C^\text{eff}_{Z'Z'}\right) \frac{g_B^2}{(4\pi)^2} \frac{a}{f_a} Z'_{\mu\nu} \tilde{Z}'^{\mu\nu}  + \left(C^\text{SM}_{Z'Z}+C^\text{eff}_{Z'Z}\right) \frac{g_B e}{s_W c_W} \frac{1}{(4\pi)^2} \frac{a}{f_a} Z'_{\mu\nu} \tilde{Z}^{\mu\nu} \nonumber \\
        & + \left(C^\text{SM}_{WW}+C^\text{eff}_{WW}\right) \frac{g_L^2}{(4\pi)^2} \frac{a}{f_a} W_{\mu\nu} \tilde{W}^{\mu\nu} + C^\text{SM}_{gg} \frac{g_s^2}{(4\pi)^2} \frac{a}{f_a} G^a_{\mu\nu} \tilde{G}^{a\mu\nu} \nonumber \\
        & + i \frac{a}{f_a} \frac{e}{\sqrt{2} s_W} (W_\mu^- (X_d J^{+\mu}_{W} - X_u J^{+\mu}_{W,\slashed{l}}) + \text{h.c.})   \ .
\label{eqn:axL_B}
\end{align}
As mentioned previously, we neglected the effects from the heavy Higgses, $h'$, $H_0$, $H'_0$ and $A_0$.  The current $J^{+\mu}_{W,\slashed{l}}$ denotes the $W$ boson current coupling without leptons. We note that the last term is generic for DFSZ models, since the couplings to the $W$ bosons do not generally commute with the PQ charges assigned to the respective weak isospin components in the quark and lepton sectors.

\section{Explicit Calculation of Axion Wilson Coefficients}
\label{sec:WilsonCoefficients}

We have built a complete set of axion and $Z'$ EFT operators in~\eqnref{axL_B} generated after integrating out the anomalons.  We remark that we can also generate a Wess-Zumino term following the analysis of Ref.~\cite{Michaels:2020fzj}. We calculate the matching conditions for the Wilson coefficients of each set of operators.  We emphasize that the final observables that are derived in this section for the Lagrangian in~\eqnref{axL_B} are basis independent in regards to chiral transformations of anomalons by construction.

\subsection{Loop-induced axion coupling to gauge bosons}
\label{subsec:avv}

\begin{figure}[htb] 
  \centering
  \begin{subfigure}{0.3\textwidth}
     \includegraphics[width=0.75\textwidth]{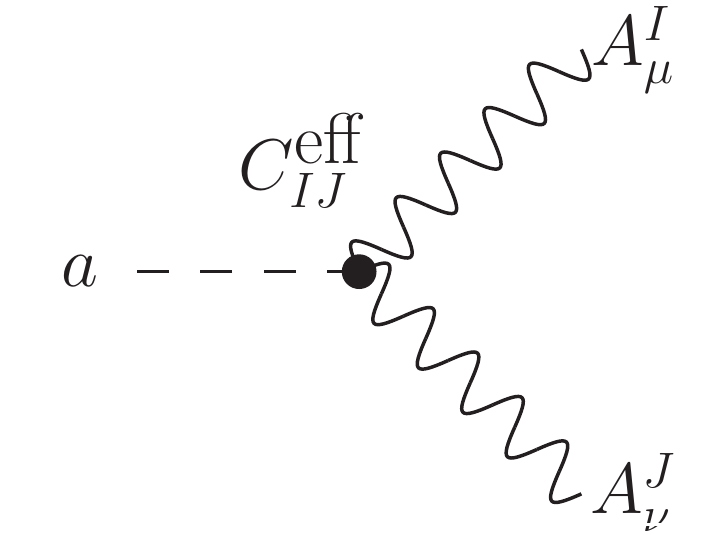}
  \end{subfigure}
  \hfill
  \begin{subfigure}{0.3\textwidth}
     \includegraphics[width=\textwidth]{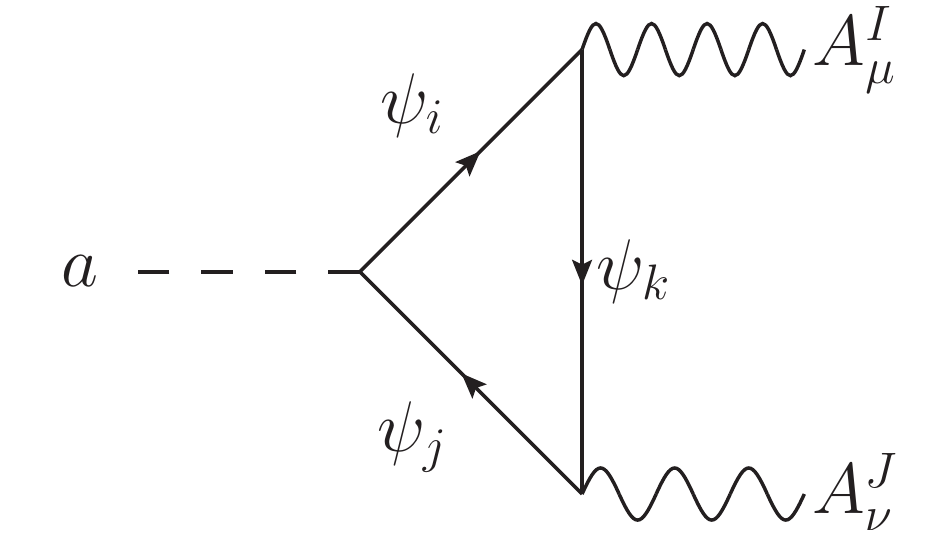}
  \end{subfigure}
  \hfill
  \begin{subfigure}{0.3\textwidth}
     \includegraphics[width=\textwidth]{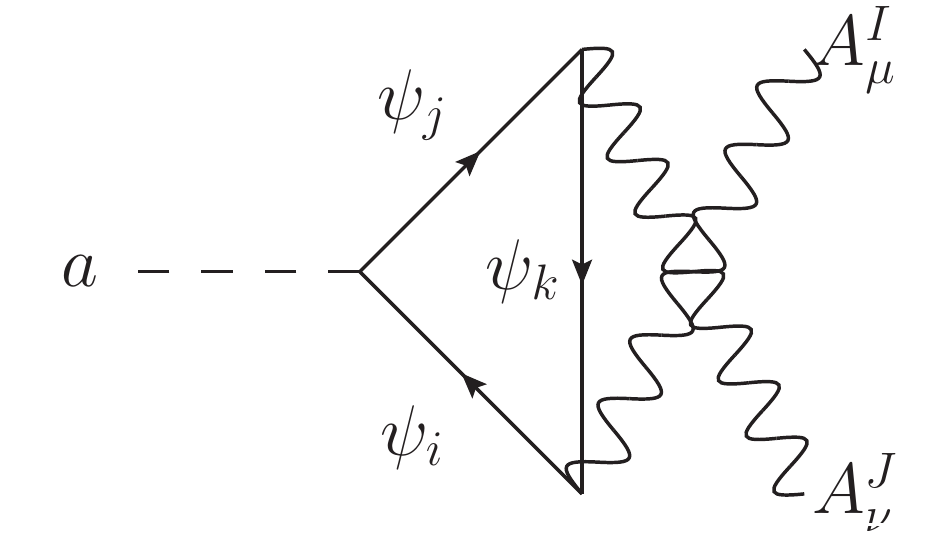}
  \end{subfigure}
  \caption[One-loop diagrams for anomalous coupling of an axion to two gauge bosons]{Effective vertex and one-loop diagrams for anomalous coupling of an axion to two gauge bosons. The loop consists of three fermions $\psi_i$, $\psi_j$ and $\psi_k$ connecting an axion $a$ with two gauge bosons $A_\mu^I$ and $A_\nu^J$.}
\label{fig:avv}
\end{figure}

We begin with the axion coupling to two gauge bosons, $C_{IJ}^\text{eff}$ for gauge bosons $A_I^\mu$ and $A_J^\mu$, where the operator is given by
\begin{equation}
\mathcal{L} \supset - C_{IJ}^\text{eff} \frac{g_I g_J}{(4\pi)^2}   
  \frac{a}{f_a} F_{I\mu\nu} \tilde{F}_J^{\mu\nu} =  -\frac{C_{IJ}^\text{eff}}{2} \frac{g_Ig_J}{(4\pi)^2} \frac{a}{f_a} \epsilon^{\mu\nu\alpha\beta} (\partial_\mu A_{I\nu} - \partial_\nu A_{I\mu}) (\partial_\alpha A_{J\beta} - \partial_\beta A_{J\alpha}) \ .
\label{eqn:Lcavv}
\end{equation}
We calculate the Wilson coefficient as a one-loop triangle diagram mediated by fermions, $\psi_i$, $\psi_j$ and $\psi_k$, as shown in~\figref{avv}.  Our calculation is performed in a general structure to allow for flavor-violating gauge interactions with both vector and axial-vector couplings.  We also introduce in general a mass to the gauge bosons and calculate in unitary gauge. The couplings of the axion are determined by the general basis at order $1/f_a$ given by~\eqnref{ALP_coup_flavor}.  

We will consider three simplifying scenarios for $C_{IJ}^\text{eff}$.  The first is the heavy anomalon limit, with $m_i,\ m_j,\ m_k \gg m_a,\ m_I,\ m_J$.  The second is the flavor-conserving limit, relevant for intermediate SM fermions.  The third is for the axion-$W \tilde{W}$ calculation, where the axion coupling is taken flavor-conserving and the $W$ couplings are necessarily left-handed.

For the heavy anomalon limit, the Wilson coefficient reads
\begin{align}
    C_{IJ}^\text{eff}=&\biggl(C_3^{IJ}-\sum_{i,j,k}N_k\frac{2C_0(0,0,0,m_i,m_j,m_k)}{\lambda(m_a^2,m_I^2,m_J^2)}\times\nonumber\\
    &\times\biggl((m_i+m_j)(C_{1A}^{ij}+C_{2A}^{ij})\biggl((Q_{IV}^{ik}Q_{JV}^{kj}-Q_{IA}^{ik}Q_{JA}^{kj})m_km_a^2(m_a^2-m_I^2-m_J^2)\nonumber\\
    &\quad +(Q_{IV}^{ik}Q_{JV}^{kj}+Q_{IA}^{ik}Q_{JA}^{kj})(m_im_J^2(m_a^2+m_I^2-m_J^2)+m_jm_I^2(m_a^2-m_I^2+m_J^2))\biggr)\nonumber\\
    &\quad+(m_i-m_j)(C_{1V}^{ij}+C_{2V}^{ij})\biggl((Q_{IA}^{ik}Q_{JV}^{kj}-Q_{IV}^{ik}Q_{JA}^{kj})m_km_a^2(m_a^2-m_I^2-m_J^2)\nonumber\\
    &\quad +(Q_{IV}^{ik}Q_{JA}^{kj}+Q_{IA}^{ik}Q_{JV}^{kj})(m_im_J^2(m_a^2+m_I^2-m_J^2)-m_jm_I^2(m_a^2-m_I^2+m_J^2))\biggr)\biggr)\nonumber\\
    &+\sum_{i,j,k}N_k\biggl(C_{1A}^{ij}(Q_{IV}^{ik}Q_{JV}^{kj}+Q_{IA}^{ik}Q_{JA}^{kj})+C_{1V}^{ij}(Q_{IV}^{ik}Q_{JA}^{kj}+Q_{IA}^{ik}Q_{JV}^{kj})\biggr)\biggr)+\mathcal{O}\biggl(\frac{m_{a,I,J}^2}{m_{i,k,j}^2} \biggr) \ ,
\label{eqn:Ceff_avv}
\end{align}
where we have expanded in ratios of the external boson mass over the internal fermion mass squared.  We remark that the axion couplings $C_1$, $C_2$ and $C_3$ coefficients from the general basis defined in~\eqnref{ALP_coup_flavor} are all included, which is necessary for ensuring the result is invariant under chiral basis transformations following~\eqnref{trafo_changes}.  The loop function is given by the three-point Passarino-Veltman function $C_0$~\cite{Passarino:1978jh}, which is given in the heavy fermion limit by
\begin{align}
    &C_0(m_{b_1}^2,m_{b_2}^2,m_{b_3}^2,m_{f_1},m_{f_2},m_{f_3})=C_0(0,0,0,m_{f_1},m_{f_2},m_{f_3})+\frac{1}{m_{f_2}^2}\mathcal{O}\biggl(\frac{m_{b_1,b_2,b_3}^2}{m_{f_1,f_2,f_3}^2}\biggr) \nonumber\\
    &=\frac{1}{m_{f_2}^2}\biggl(\frac{\ln(m_{f_1}^2/m_{f_2}^2)}{(1-m_{f_3}^2/m_{f_1}^2)(1-m_{f_1}^2/m_{f_2}^2)}+\frac{\ln(m_{f_3}^2/m_{f_2}^2)}{(1-m_{f_1}^2/m_{f_3}^2)(1-m_{f_3}^2/m_{f_2}^2)}+\mathcal{O}\biggl(\frac{m_{b_1,b_2,b_3}^2}{m_{f_1,f_2,f_3}^2}\biggr)\biggr) \ .
\label{eqn:C0_exp}
\end{align}
The explicit Wilson coefficient for an axion coupling to two $SU(N)$ gauge bosons is found by replacing $N_k \to T(R_{Ik})$. The matrix element vanishes for $I \neq J$ in the case of an $SU(N)$ gauge group. 

In the flavor-conserving limit, $i = j = k \equiv f$, where the axion-fermion Yukawa interactions are only given by $C_{2A}^f$, we have
\begin{align}
\left( C_{IJ}^\text{eff} \right)_f = & -4 \sum_fN_fC_{2A}^f\biggl(Q_{IV}^fQ_{JV}^fm_f^2C_0(m_a^2,m_I^2,m_J^2,m_f,m_f,m_f)\nonumber\\
    &-\frac{Q_{IA}^fQ_{JA}^fm_f^2}{\lambda(m_a^2,m_I^2,m_J^2)}\biggl((m_a^2+m_I^2-m_J^2)(m_a^2-m_I^2+m_J^2)C_0(m_a^2,m_I^2,m_J^2,m_f,m_f,m_f)\nonumber\\
    &\hspace{3.5cm}+4m_a^2B_0(m_a^2,m_f,m_f)-2(m_a^2+m_I^2-m_J^2)B_0(m_I^2,m_f,m_f)\nonumber\\
    &\hspace{3.5cm}-2(m_a^2-m_I^2+m_J^2)B_0(m_J^2,m_f,m_f)\biggr)\biggr) \ .
\label{eqn:CeffSM1}
\end{align}
Here, $B_0$ is the standard Passarino-Veltman two-point scalar function~\cite{Passarino:1978jh}, whose divergence exactly cancels.  The Wilson coefficient $C_\text{eff}^{gg}$ for gluons is again given by replacing $N_f \to T(R_{If})$. 

Lastly, we consider the $W_\mu^\pm$ gauge bosons, which are not covered by the flavor-conserving limit, but we still assume the axion coupling is flavor-conserving, $i = j \equiv f$.  The corresponding Wilson coefficient is then given by 
\begin{align}
    C_{WW}^\text{eff}=&4\sum_f N_f C_{2A}^f\frac{Q_{WL}^{fk}Q_{WL}^{kf}m_f^2}{m_a^2-4m_W^2}\biggl(B_0(m_a^2,m_f,m_f)\nonumber\\
    &-B_0(m_W^2,m_f,m_k)+(m_W^2-m_f^2+m_k^2)C_0(m_a^2,m_W^2,m_W^2,m_f,m_f,m_k)\biggr) \ ,
\label{eqn:CeffSM2}
\end{align}
where the divergences from $B_0$ again cancel.  These two limiting cases in~\eqnsref{CeffSM1}{CeffSM2} will be used to evaluate the contributions to axion couplings to gauge bosons arising from SM fermions.

\subsection{Loop-induced axion coupling to scalar and gauge boson}
\label{subsec:avs}

\begin{figure}[htb] 
  \centering
  \begin{subfigure}{0.3\textwidth}
     \includegraphics[width=0.75\textwidth]{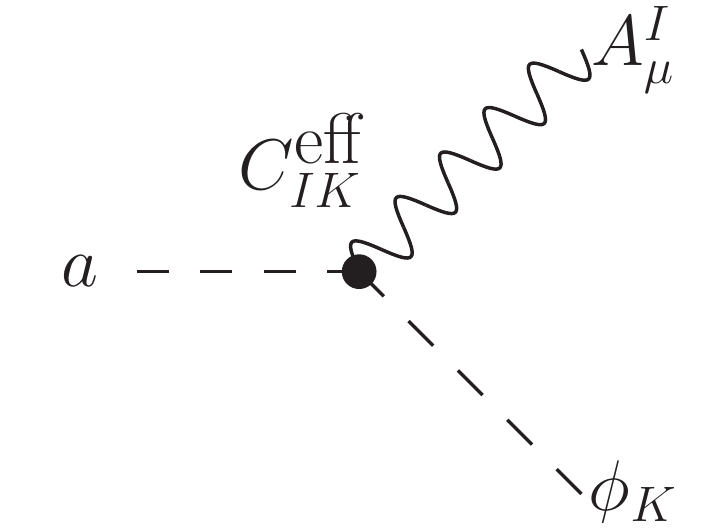}
     \caption{}
     \label{fig:avs0}
  \end{subfigure}
  \hfill
  \begin{subfigure}{0.3\textwidth}
     \includegraphics[width=\textwidth]{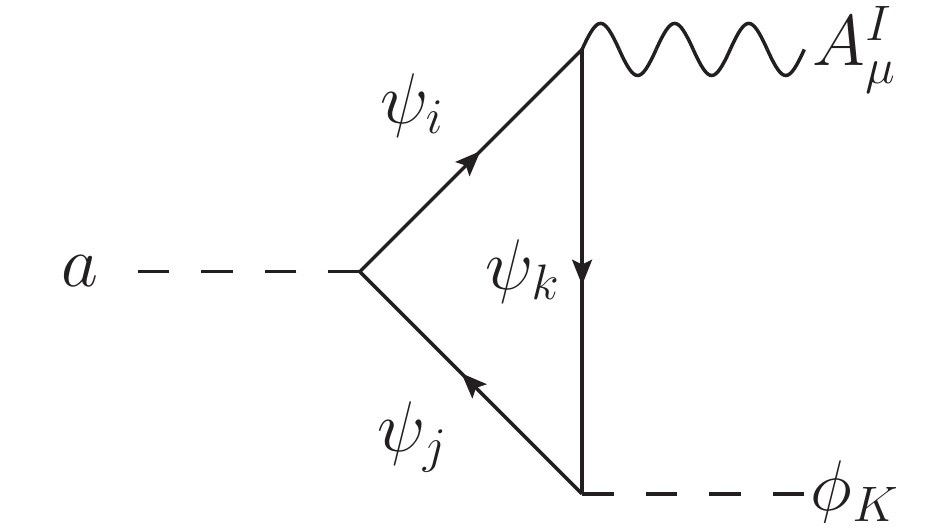}
     \caption{}
     \label{fig:avs1}
  \end{subfigure}
  \hfill
  \begin{subfigure}{0.3\textwidth}
     \includegraphics[width=\textwidth]{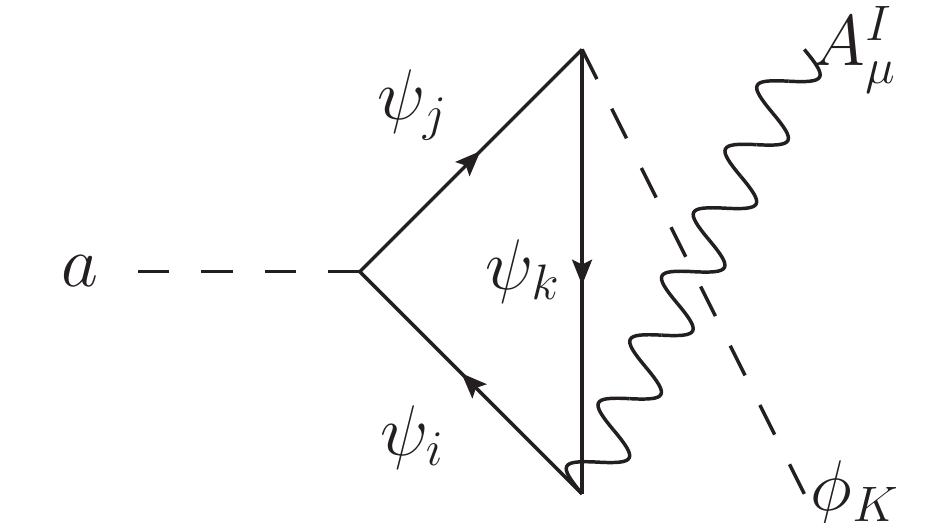}
     \caption{}
     \label{fig:avs2}
  \end{subfigure}
  \vskip\baselineskip
  \begin{subfigure}{0.3\textwidth}
     \includegraphics[width=0.75\textwidth]{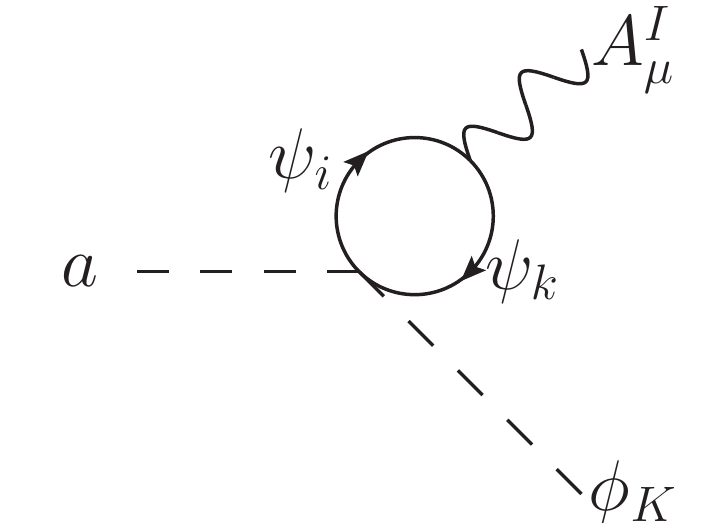}
     \caption{}
     \label{fig:avs3}
  \end{subfigure}
  \hfill
  \begin{subfigure}{0.3\textwidth}
     \includegraphics[width=0.75\textwidth]{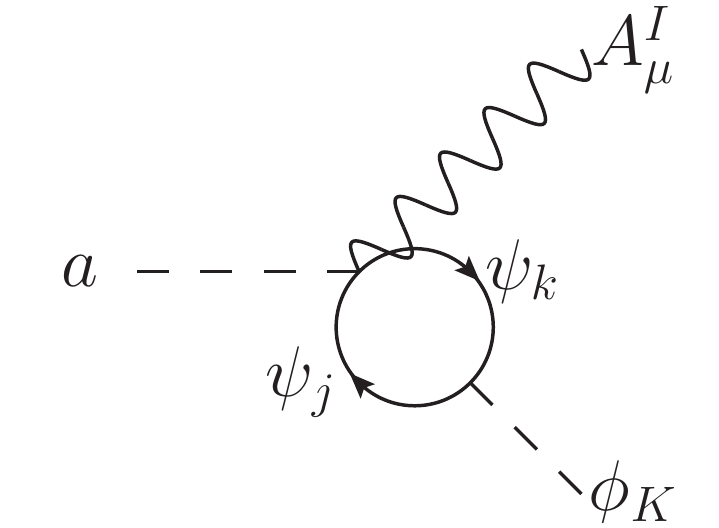}
     \caption{}
     \label{fig:avs4}
  \end{subfigure}
  \hfill
  \begin{subfigure}{0.3\textwidth}
     \includegraphics[width=\textwidth]{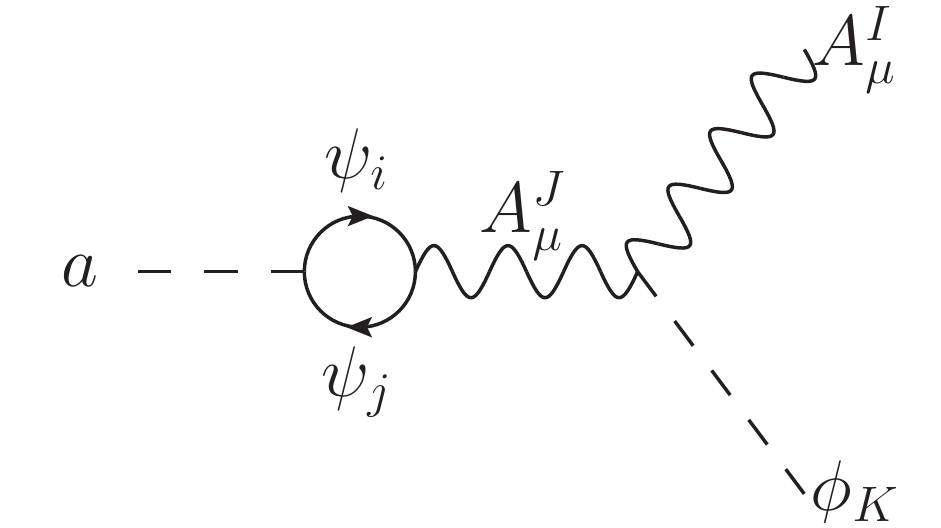}
     \caption{}
     \label{fig:avs5}
  \end{subfigure}
  \caption[One-loop diagrams for coupling of an axion to a gauge boson and a scalar]{Effective vertex and one-loop diagrams for coupling of an axion to a gauge boson and a scalar. The loop consists of three fermions $\psi_i$, $\psi_j$ and $\psi_k$ connecting an axion $a$ with one gauge boson $A_\mu^I$ and a scalar $\phi_K$.}
  \label{fig:avs}
\end{figure}

In this subsection, we calculate the Wilson coefficient of an axion coupling to a gauge boson $A_\mu^I$ and a scalar $\phi_K$ at one loop from integrating out the anomlons under the use of the heavy fermion limit. The corresponding operator of dimension four is defined by
\begin{equation}
    \mathcal{L}\supset-C_{IK}^\text{eff}\frac{g_I}{(4\pi)^2}\phi_KA_\mu^I\partial^\mu a \ ,
\end{equation}
and is shown schematically in~\figref{avs0}.  The one-loop diagram is the leading contribution for the interaction between the axion, the SM Higgs $h$ and the gauge bosons $Z_\mu$ and $Z'_\mu$, but there can be other tree-level contributions if the SM Higgs has a mass mixing to other scalar fields. 

Besides the usual triangle diagram, which is shown in~\figsref{avs1}{avs2}, there are three more non-vanishing contributions at one-loop order.  The diagrams in~\figsref{avs3}{avs4} are induced by the five-dimensional contact interactions from the general Lagrangian in~\eqnref{ALP_coup_flavor}. In addition, the diagram in~\figref{avs5} consists of a mixing of the axion into an internal off-shell gauge boson propagator which then couples to the Higgs: this kinetic mixing vanishes for $A_\mu^J$ being on-shell.
We remark that this effective interaction has also been studied in the context of Ref.~\cite{Bauer:2016zfj} as an bottom-up discriminator of the coupling origin of the pseudoscalar SM singlet to the Higgs and $Z$ boson.  In comparison to Ref.~\cite{Bauer:2016zfj}, we must include one additional diagram in~\figref{avs4} because of the commutator from~\eqnref{comm_int}.

Now, we calculate the effective coefficient in the heavy fermion limit. For the diagram in~\figref{avs5}, we define the coupling of a scalar to gauge bosons as
\begin{equation}
\mathcal{L} \supset \frac{1}{2}(\phi_K+v_K)^2 \biggl( \sum_I g_I Q_I^K A^I_{\mu} \biggr) \biggl( \sum_J g_J Q_J^K A^{J\mu} \biggr) \ .
\end{equation}
We see that the coupling linear in $\phi_K$ is proportional to $v_K$, and thus the gauge bosons must be massive.  In particular, scalar fields orthogonal to a vev do not induce this diagram in the Higgs basis.

The effective coefficient then reads
\begin{align}\label{eqn:Ceff_asv}
    C_{IK}^\text{eff}&=\frac{1}{(4\pi)^2}\frac{1}{f_a}\frac{1}{\sqrt{2}}\sum_{i,j,k}N_k\biggl(2\frac{C_0(0,0,0,m_i,m_j,m_k)}{\lambda(m_a^2,m_K^2,m_I^2)} \nonumber\\
    &\times(m_i+m_j)((C_{1A}^{ji}+C_{2A}^{ji})Q_{IA}^{ik}Y_K^{kj}+(C_{1A}^{ij}+C_{2A}^{ij})Y_K^{jk}Q_{IA}^{ki}) \nonumber\\
    &\times\biggl(m_I^2(m_I^2-m_a^2-m_K^2)(m_i-m_j)(m_j+m_k)-\lambda(m_a^2,m_K^2,m_I^2)(m_im_k+m_j^2)\biggr)\nonumber\\
    &-2(m_i+m_j)((C_{1A}^{ji}+C_{2A}^{ji})Q_{IA}^{ik}Y_K^{kj}+(C_{1A}^{ij}+C_{2A}^{ij})Y_K^{jk}Q_{IA}^{ki})B_0(0,m_k,m_i)\nonumber\\
    &+(m_j-m_k)((C_{1A}^{ji}+C_{IA}^{ji})Q_{IA}^{ik}Y_K^{kj}+(C_{1A}^{ij}+C_{IA}^{ij})Y_K^{jk}Q_{IA}^{ki})B_0(0,m_j,m_k)\nonumber\\
    &-\sum_{J,L}\frac{g_J^2Q_I^KQ_J^Kv_Kv_L}{m_J^2}(m_i+m_j)((C_{1A}^{ji}+C_{2A}^{ji})Q_{JA}^{ik}Y_L^{kj}+(C_{1A}^{ij}+C_{2A}^{ij})Y_L^{jk}Q_{JA}^{ki})B_0(0,m_i,m_j)\biggr)\nonumber\\
    &+\mathcal{O}\biggl(\frac{m_{a,I,K}^2}{m_{i,k,j}^2}\biggr) \ .
\end{align}
For this result, we needed to take the heavy fermion limit of the $B_0$ function given by
\begin{align}
    &B_0(m_b^2,m_{f_1},m_{f_2})=B_0(0,m_{f_1},m_{f_2})+\mathcal{O}\left(\frac{m_b^2}{m_{f_1}^2,m_{f_2}^2}\right)\nonumber\\
    &=\frac{1}{\epsilon}-\gamma_E+\ln(4\pi)+\ln(\frac{\mu^2}{m_{f_1}m_{f_2}})+1-\frac{1}{2}\frac{m_{f_1}^2+m_{f_2}^2}{m_{f_1}^2-m_{f_2}^2}\ln(\frac{m_{f_1}^2}{m_{f_2}^2})+\mathcal{O}\left(\frac{m_b^2}{m_{f_1}^2,m_{f_2}^2}\right) \ .
\end{align}
In this case the divergence of $B_0$ does not cancel trivially, but for our case of interest, we have a vanishing divergence.  For example, extracting the divergent term for the color-singlet case, $N_k = 1$, we get 
\begin{align}
    C_{IK}^\text{eff}=&\frac{1}{(4\pi)^2}\frac{1}{f_a}\frac{1}{\sqrt{2}}\frac{1}{\epsilon}\biggl(\mathcal{T}_-\biggl(\mathbf{Y}_K,\mathbf{M},\mathbf{Q}_{IA},\mathbf{C}_{2A}-\mathbf{C}_{IA}\biggr)+\mathcal{T}_+\biggl(\mathbf{Y}_K,\mathbf{M},\mathbf{Q}_{IA},\mathbf{C}_{1A}+\mathbf{C}_{2A}\biggr)\nonumber\\
    &-\sum_{J,L}\frac{g_J^2Q_I^KQ_J^Kv_Kv_L}{m_J^2}\mathcal{T}_+\biggl(\mathbf{Y}_L,\mathbf{M},\mathbf{Q}_{JA},\mathbf{C}_{1A}+\mathbf{C}_{2A}\biggr)\biggr)+\mathcal{O}(\epsilon^0) \ ,
\end{align}
where we used the notation
\begin{align}
    \mathcal{T}_+(\mathbf{A},\mathbf{B},\mathbf{C},\mathbf{D})&=\Tr[\{\mathbf{A},\mathbf{C}\}\{\mathbf{B},\mathbf{D}\}]+\Tr[\{\mathbf{B},\mathbf{C}\}\{\mathbf{A},\mathbf{D}\}] \ , \\
    \mathcal{T}_-(\mathbf{A},\mathbf{B},\mathbf{C},\mathbf{D})&=\Tr[\{\mathbf{A},\mathbf{C}\}\{\mathbf{B},\mathbf{D}\}]-\Tr[\{\mathbf{B},\mathbf{C}\}\{\mathbf{A},\mathbf{D}\}]=\Tr[[\mathbf{B},\mathbf{A}][\mathbf{C},\mathbf{D}]] \ .
\end{align}
In the Higgs basis, we have the relation
\begin{equation}
    \frac{g_J Q_I^K v_L}{m_J} = \delta_{IJ}\delta_{KL} \ ,
\end{equation}
such that the divergent term simplifies to
\begin{align}
    C_{IK}^\text{eff}=&\frac{1}{(4\pi)^2}\frac{1}{f_a}\frac{1}{\epsilon}\mathcal{T}_-\biggl(\mathbf{Y}_K,\mathbf{M},\mathbf{Q}_{IA},\mathbf{C}_{2A}-\mathbf{C}_{IA}\biggr)+\mathcal{O}(\epsilon^0) \ .
\end{align}
For the anomalons, the commutator $[\mathbf{M},\mathbf{Y}_K]$ vanishes and thus the divergence cancels.

Again, the result in~\eqnref{Ceff_asv} is invariant under chiral basis transformations, as seen using the transformation properties in~\eqnref{trafo_changes}.  This check requires the inclusion of the diagram in~\figref{avs4} in the flavor-violating case.  We also observe that the Wilson coefficient only depends on axial gauge charges, leading to couplings for $Z$ and $Z'$ bosons but not the photon.

\subsection{Wilson coefficients and parameters in the DFSZ model with gauged baryon number}
\label{subsec:WilsonDFSZgB}

Now we determine the Wilson coefficients from integrating out the heavy anomalons $E_{1,2}$ and $N_{1,2}$ in our model. Afterwards we identify the parameter space for our model. The coefficients of the ALP-EFT are defined in~\eqnref{axL_B}. We use the heavy fermion expansion for the coefficients defined in~\eqnref{Ceff_avv} and~\eqnref{Ceff_asv} and drop all terms of order $1/f_a$, $\epsilon_\text{eff}^2$, $\Delta_E^2$, $\Delta_N^2$ and $\Delta_{EN}^2$, if not mentioned otherwise. We obtain
\begin{align}
C_{\gamma\gamma}^\text{eff} = &-\frac{8}{3}Xs_{\beta '}^2\frac{\Delta_E}{\Sigma_E^3}\cos(2\alpha_E)\frac{m_a^2}{f_a^2}+\mathcal{O}\left(\frac{1}{f_a^3}\right) \ , \label{eqn:CGammaGamma}\\
C_{\gamma Z}^\text{eff} = &-\frac{Xs_{\beta '}^2}{4} \ ,\quad C_{ZZ}^\text{eff} = -\frac{Xs_{\beta '}^2}{4}(1-2s_W^2) \ , \quad C_{WW}^\text{eff} = -\frac{Xs_{\beta '}^2}{2} \ , \label{eqn:CZW}\\
C_{hZ'}^\text{eff} = &-\frac{Xs_{\beta '}^2}{2}\frac{v}{f_a}(\Sigma_M^2+\Delta_M^2)\left(1-6\frac{m_{Z'}^2(m_a^2+m_h^2-m_{Z'}^2)}{\lambda(m_{Z'}^2,m_a^2,m_h^2)}\right)+\mathcal{O}\left(\frac{1}{f_a^2}\right) \ , \label{eqn:ChZp} \\
C_{\gamma Z'}^\text{eff} =& -\frac{\epsilon_\text{eff}e}{g_Bc_W}\frac{m_{Z'}^2}{m_{Z'}^2-m_{Z}^2}C_{\gamma Z}^\text{eff} \ ,  \quad 
 C_{hZ}^\text{eff} = -\frac{g_B\epsilon_\text{eff}s_W^2c_W}{e}\frac{m_{Z}^2}{m_{Z}^2-m_{Z'}^2}C_{hZ'}^\text{eff} \ , \\ \quad C_{ZZ'}^\text{eff} =& -\frac{\epsilon_\text{eff}e}{g_Bc_W}\frac{m_{Z'}^2}{m_{Z'}^2-m_{Z}^2}C_{ZZ}^\text{eff} \ , \quad
C_{Z'Z'}^\text{eff} = \left(-\frac{\epsilon_\text{eff}e}{g_Bc_W}\frac{m_{Z'}^2}{m_{Z'}^2-m_{Z}^2}\right)^2C_{ZZ}^\text{eff}+\mathcal{O}\left(\epsilon_\text{eff}^3\right) \  . 
\end{align}
We discarded the interactions to the new scalars $h'$, $H_0$, $H'_0$ and $A_0$ since their masses can naturally be taken to be larger than the spectrum of interest.  In these expressions, we introduce the parameters denoted by $\Sigma$ ($\Delta$) to describe mass sums (differences) defined as
\begin{align}
    \Sigma_M &= \frac{m_{12}+m_{34}}{v}=\frac{1}{2}\left(\frac{|y_1|+|y_2|}{\sqrt{2}}c_{12}c_\beta+\frac{|y_3|+|y_4|}{\sqrt{2}}c_{34}s_\beta\right)\approx\frac{c_\beta+s_\beta}{\sqrt{2}} \ , \\
    \Delta_M &= \frac{m_{12}-m_{34}}{v}=\frac{1}{2}\left(\frac{|y_1|+|y_2|}{\sqrt{2}}c_{12}c_\beta-\frac{|y_3|+|y_4|}{\sqrt{2}}c_{34}s_\beta\right)\approx\frac{c_\beta-s_\beta}{\sqrt{2}} \ , \\
    \Sigma_E &= \frac{m_{E_1}+m_{E_2}}{f_a}=\frac{2m_{LE}}{f_a}=\frac{|y_L|+|y_E|}{\sqrt{2}}\frac{c_\gamma}{Xs_{\beta'}}\approx \frac{4\pi}{3}\frac{\sqrt{2}}{Xs_{\beta'}} \ , \\
    \Sigma_N &= \frac{m_{N_1}+m_{N_2}}{f_a}=\frac{2m_{LN}}{f_a}=\frac{|y_L|+|y_N|}{\sqrt{2}}\frac{c_\gamma}{Xs_{\beta'}}\approx \frac{4\pi}{3}\frac{\sqrt{2}}{Xs_{\beta'}} \ , \\
    \Delta_{E} &= \frac{m_{E_1}-m_{E_2}}{f_a}=-\frac{1}{\sin(2\alpha_E)}\frac{2m_{12}}{f_a}=-\frac{v}{f_a}\frac{\Sigma_M+\Delta_M}{\sin(2\alpha_E)}\approx-\frac{v}{f_a}\frac{\sqrt{2}c_\beta}{\sin(2\alpha_E)} \ , \\
    \Delta_{N} &= \frac{m_{N_1}-m_{N_2}}{f_a}=-\frac{1}{\sin(2\alpha_N)}\frac{2m_{34}}{f_a}=-\frac{v}{f_a}\frac{\Sigma_M-\Delta_M}{\sin(2\alpha_N)}\approx-\frac{v}{f_a}\frac{\sqrt{2}s_\beta}{\sin(2\alpha_N)} \ , \\
    \Delta_{EN} &= \Sigma_E-\Sigma_N=\frac{2m_{LE}-2m_{LN}}{f_a}=\frac{|y_E|-|y_N|}{\sqrt{2}}\frac{c_\gamma}{Xs_{\beta'}} \ .
\end{align}
For the above approximations, we assumed that the Yukawa couplings $|y_L|$, $|y_E|$ and $|y_N|$ are $4\pi/3$ and the other Yukawa couplings are of order 1. In addition, we assumed that we can use the small angle approximation for the angles $\gamma$, $\delta_{12}$ and $\delta_{34}$. 

We highlight that the axion diphoton coupling is significantly suppressed, as a result of the fact that the flavor-conserving interactions of the axion to the charged anomalons come with opposite signs, and so the induced remainder is proportional to the mass difference of the charged anomalons.  The same would be true for the axion coupling to two $Z'$ bosons, but the coupling shifts owing to the kinetic mixing are important.  Hence, in contrast to standard axion EFTs where the diphoton and the digluon coupling dominate, we find the coefficients $C_{\gamma Z}^\text{eff}$, $C_{ZZ}^\text{eff}$ and $C_{WW}^\text{eff}$ as the most important coefficients.  We will investigate the corresponding implications in~\subsecref{BRs}, when we consider collider-scale masses for the axion.

Although we have prescribed a DFSZ-like structure for the QCD axion, we will expand the model to include the possibility of non-QCD instanton contributions to the axion mass.  For this purpose, we will distinguish a QCD axion and an ALP via their topological susceptibility $\chi$ defined by~\eqnref{topsus}, which fixes the relationship between $m_a$ and $f_a$.

The $Z'$ mass and the anomalon masses also implicitly depend on $f_a$ since in the invisible axion limit, $f_a$ is mainly composed of $v'$ as shown in~\subsecref{scalars} with $f_a = X c_\gamma^{-1}v' s_{\beta '}c_{\beta '}$. The corresponding relations are
\begin{equation}
    m_{Z'} = 3 g_B v' = \frac{3 c_\gamma g_B f_a}{X s_{\beta'} c_{\beta'}} \ , \quad 
    m_{\text{anom}} \equiv \frac{\Sigma_{E,N} \pm \Delta_{E,N}}{2} f_a \approx \frac{4\pi}{3}
    \frac{f_a}{X s_{\beta '} \sqrt{2}} \ .
\label{eqn:vprime_to_fa}
\end{equation}
Thus, to have a $m_{Z'}$ which is smaller than the mass scale of the anomalons but allows a sizeable gauge coupling $g_B$ we can assume the case of maximal mixing between the additional scalar fields ($\beta'=\pi/4$). 

As a last detail, we normalize our results analogously to the canonical form of the $G\tilde{G}$ operator in the literature by rescaling with the PQ charge normalization $X$. Namely, we write $X \equiv \frac{1}{2\mathcal{A}_{gg}}$ where $\mathcal{A}_{gg}$ describes the color anomaly defined by
\begin{equation}
    \mathcal{L}\supset \frac{g_s^2}{(4\pi)^2}X\mathcal{A}_{gg}\frac{a}{f_a}G_{\mu\nu}^a\tilde{G}^{a\mu\nu}=\frac{g_s^2}{32\pi^2}\frac{a}{f_a}G_{\mu\nu}^a\tilde{G}^{a\mu\nu}.
\end{equation}
In our model, where only the SM fermions couple to the gluons, we get $\mathcal{A}_{gg} = 3$ and thus $X = \frac{1}{6}$.  Hence, our parameters simplify to $m_{Z'} \approx 36 g_B f_a$ and $m_\text{anom} \approx 8\pi f_a$. This leaves us with two independent parameters for a model with a QCD axion, $g_B$ and $f_a$, as well as a third parameter $\chi$ for an ALP. We can also trade $f_a$ or $\chi$ in favor of having $m_a$ as a free parameter as well as $f_a$ or $g_B$ in favor of having a free $m_{Z'}$.

Now that we have determined the exact form of the Wilson coefficients from the complete set of operators, presented in the Lagrangian in~\eqnref{axL_B}, we can discuss the varied phenomenological implications of this model.

\section{Axion and $Z'$ EFT Phenomenology}
\label{sec:pheno}

In this section, we investigate phenomenological aspects of our model, where we focus on the axion $a$ and the additional gauge boson $Z'_\mu$. We focus on collider scale masses ranging from the elctroweak scale to multi-TeV and the numerous possible resonance channels dictated by the various decays.  We first present a branching ratio analysis of $a$ and $Z'_\mu$, built from the EFT operator analysis from~\secref{EFT}.  Following this, we derive the current constraints in the $\{m_{Z'},g_B\}$ and $\{m_a,G_{a\gamma\gamma}\}$ planes for $Z'$ and $a$, respectively. 

\subsection{Branching ratios for axion/ALP and $Z'$ decays}
\label{subsec:BRs}

As noted in~\secref{EFT}, the heavy anomalon limit dramatically reduces the parameter space dependence of the axion and $Z'$ EFT, leaving a comprehensive set of EFT operators that determine the production and decay modes of these new physics particles with a definite pattern of coefficients from~\eqnref{CGammaGamma}-~\eqnref{ChZp}, for example.  Hence, to study the patterns of ALP branching ratios, we only need to specify the $Z'$ mass and $f_a$.  

\begin{figure}[htb] 
  \centering
  \includegraphics[width=1.0\textwidth]{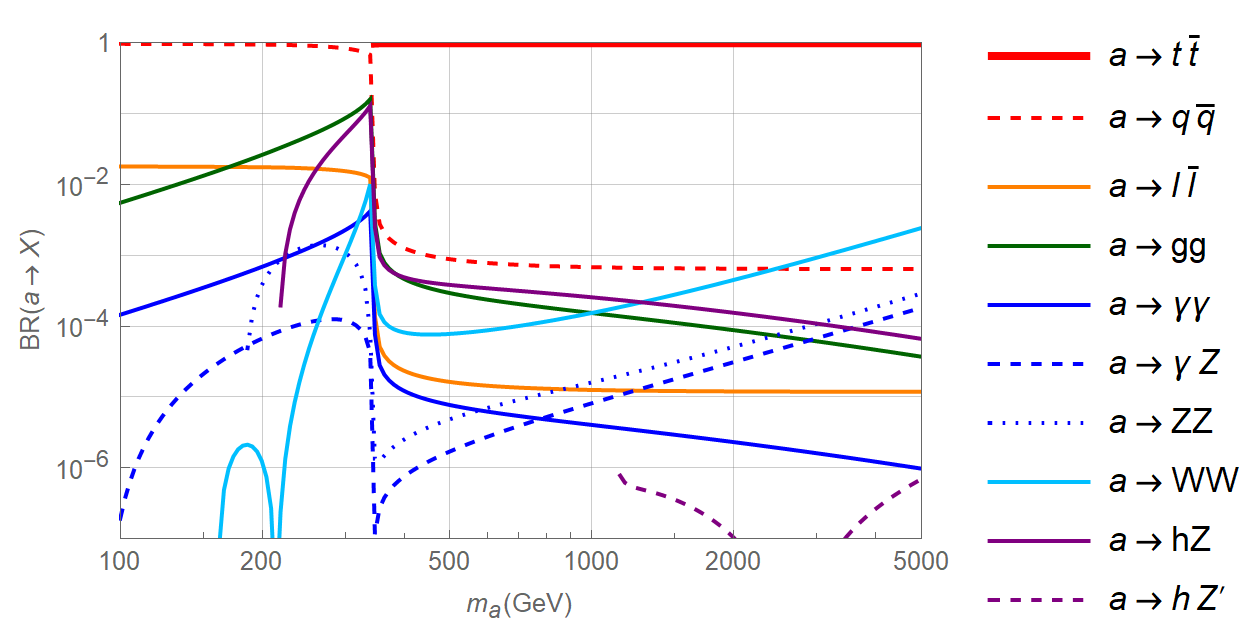}
  \caption{Branching ratios of the ALP, fixing $m_{Z'} = 1$~TeV and the ALP decay constant $f_a = 500$~GeV.}
  \label{fig:ALP_BR}
\end{figure}

In~\figref{ALP_BR}, we set the $Z'$ mass to 1~TeV and $f_a = 500$~GeV and show the ALP branching ratios as a function of its mass. The decay into two leptons sums over all charged leptons, while the decay into quarks sums over the first five quark flavors.
As evident from the figure, the most dominant decays are to quarks, if the ALP is below the top threshold, or to two top quarks, once the ALP mass increases above $2 m_t$.  This feature arises solely from the ALP coupling to the mass of the fermions.  The coupling to two gluons mirrors the coupling to two photons, reflecting the DFSZ nature of the ALP interactions and the near absence of the diphoton coupling from anomalons.  The exotic mode of $a \to hZ$ becomes enhanced near the top threshold and reaches 10\%, making it an exciting prospect for discovery.  For high ALP masses the decays into $WW$, $ZZ$ and $\gamma Z$ start to take over. These decays correspond to the anomalon mediated decays which are not suppressed as shown in~\eqnref{CZW}. Another decay which is dominantly mediated by the anomalons is $a\to hZ'$. The corresponding Wilson coefficient in~\eqnref{ChZp} changes sign at $m_a^2 = m_h^2 + 4m_{Z'}^2 + \sqrt{16 m_h^2 m_{Z'}^2 + 9 m_{Z'}^4}$, such that the branching ratio has a minimum at this ALP mass.

\begin{figure}[htb] 
  \centering
  \includegraphics[width=1.0\textwidth]{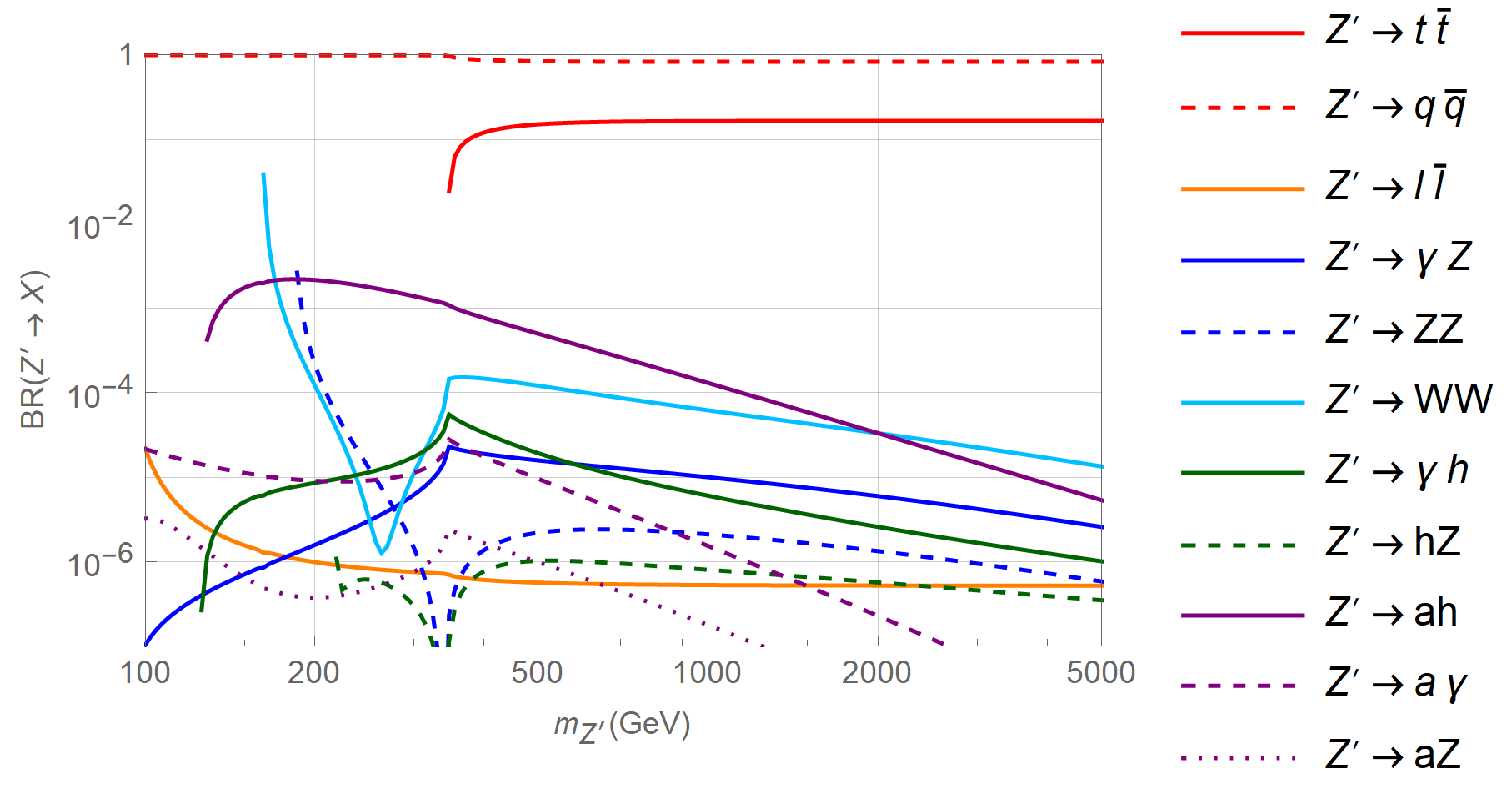}
  \caption{Branching ratios of two-body decays of the $Z'$ gauge boson for $Z'$ masses between 100 GeV and 5 TeV. The ALP mass is taken to be $m_a = 1$~GeV and the gauge coupling to be $g_B = 0.5$.}
  \label{fig:Zp_BR_fixgB}
\end{figure}

To analyze the $Z'$ branching fractions as shown in~\figref{Zp_BR_fixgB}, we fix the ALP mass $m_a = 1$~GeV and $g_B = 0.5$. 
Of course, the dominant $Z'$ decay mode is to two quarks, owing to the gauged baryon number, and $q\bar{q}$ annihilation is the main production mode for the $Z'$ at the LHC.  The corresponding decay width is given by 
\begin{equation}
    \Gamma_{Z'\to\bar{q}q}=\frac{N_q^2g_B^2B_q^2m_{Z'}}{12\pi}\sqrt{1-2\frac{m_q^2}{m_{Z'}^2}}\left(1+2\frac{m_q^2}{m_{Z'}^2}\right) \ .
\end{equation}
The decay into leptons is induced by the kinetic mixing with the $Z$ boson and is therefore suppressed by $\epsilon_\text{eff}^2$.  The decay width for the decay into leptons reads
\begin{align}
    \Gamma_{Z'\to\bar{l} l} &= \frac{e^2\epsilon_\text{eff}^2m_{Z'}}{12\pi c_W^2}\left(1-\frac{m_Z^2}{m_{Z'}^2}\right)^{-2}\biggl(\left(Q_ls_W^2\right)^2-T_3^lQ_ls_W^2+\frac{1}{2}\left(T_3^l\right)^2\biggr)\sqrt{1-2\frac{m_l^2}{m_{Z'}^2}}\nonumber\\
    &\times\biggl(1+2\frac{m_l^2}{m_{Z'}^2}\frac{(Q_ls_W^2)^2-T_3^lQ_ls_w^2-(T_3^l)^2/4}{(Q_ls_W^2)^2-T_3^lQ_ls_w^2+(T_3^l)^2/2}\biggr) \ .
\end{align}
The coupling of the leptons to the $Z$ is parameterized by the electric charge $Q_l$ and the isospin charge $T_3 = \pm 1/2$ associated to the third generator of $SU(2)_L$.

We see the exotic decay $Z' \to ah$ is relevant for light $Z'$ masses and can provide an interesting exotic production mode for the axion in a post-discovery scenario for the $Z'$ boson.  We also note the relative importance of the $WW$ channel induced by the kinetic mixing, as well as the $\gamma Z$ decay, which uses the calculation established in Ref.~\cite{Michaels:2020fzj}.  The $Z' \to Z \gamma$ exotic decay width, up to corrections of order $\mathcal{O}(\Delta_E^2,\ \Delta_N^2,\ \Delta_{EN}^2)$, is
\begin{align}
    \Gamma_{Z'\to\gamma Z}=&\frac{3}{2048 \pi^5 }\frac{e^4 g_B^2}{s_W^2 c_W^2}\frac{m_Z^2}{m_{Z'}}\left(1-\frac{m_Z^4}{m_{Z'}^4}\right)\nonumber\\
    &\left|-\sum_{q} 2T_3^q Q_q B_q\left(\frac{m_{Z'}^2}{m_{Z'}^2 - m_Z^2}\left(B_0(m_{Z'}^2, m_q, m_q)-B_0(m_{Z}^2, m_q, m_q) \right) \right.\right.\nonumber\\
    &\hspace{3cm}\left.\left. + 2m_q^2\frac{m_{Z'}^2}{m_Z^2}C_0(0,m_Z^2,m_{Z'}^2,m_q,m_q,m_q)\right)\right. \nonumber\\
    &\left.+\left(\frac{m_{Z'}^2}{m_{Z'}^2-m_Z^2}\left(B_0(m_{Z'}^2,m_\text{anom},m_\text{anom})-B_0(m_{Z}^2,m_\text{anom},m_\text{anom})\right)\right.\right.\nonumber\\
    &\hspace{3cm}+2m_\text{anom}^2C_0(0,m_Z^2,m_{Z'}^2,m_\text{anom},m_\text{anom},m_\text{anom})\biggr)\biggr|^2 \ .
\end{align}
This reflects the $Z \to Z' \gamma$ decay calculation in Ref.~\cite{Michaels:2020fzj} where the roles of the $Z'$ and $Z$ boson are interchanged.

\subsection{Collider constraints on the $Z'$ boson}
\label{subsec:colliderZp}

\begin{figure}[ht]
  \centering
    \includegraphics[width=1\textwidth]{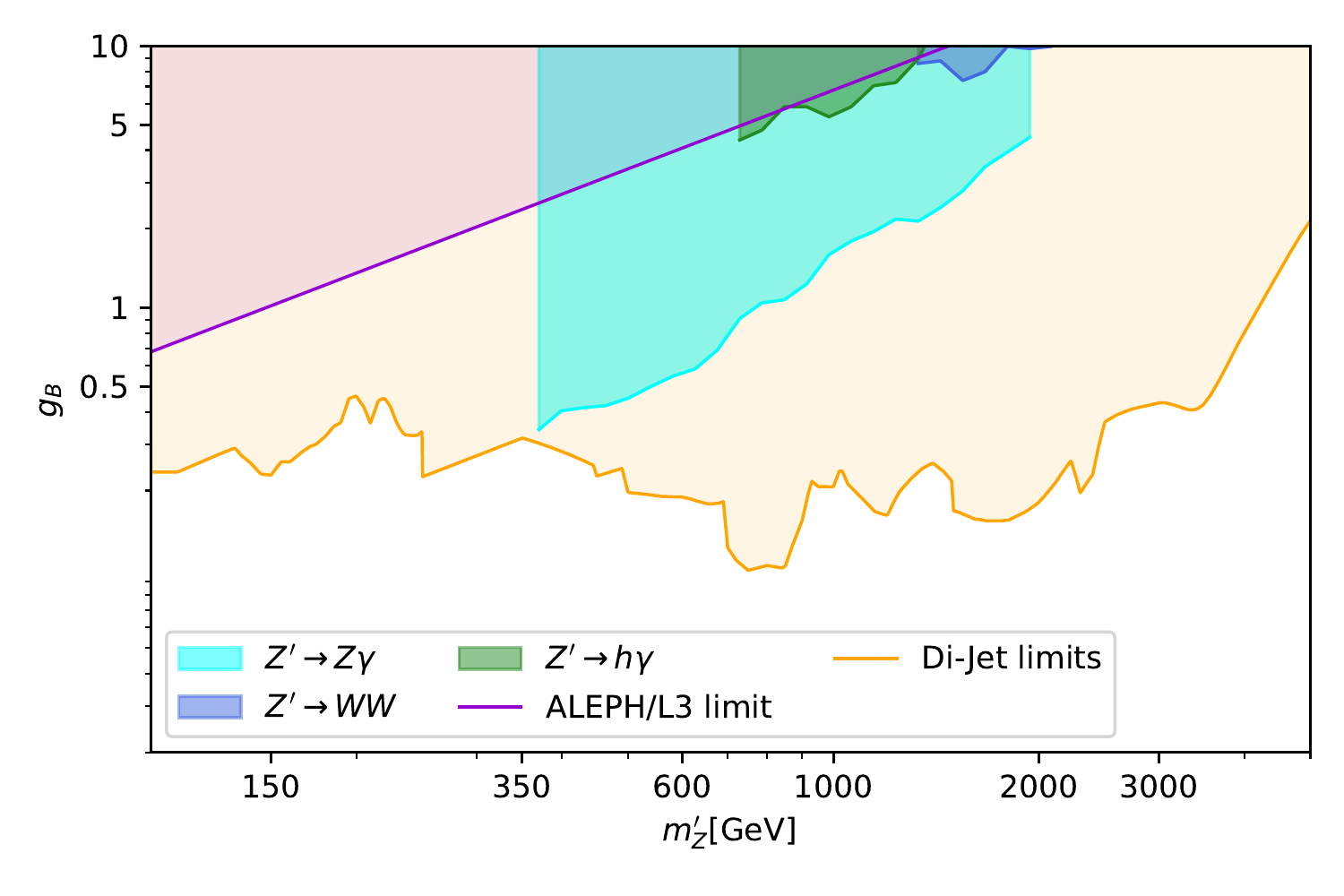}
  \caption{ 
  Constraints for the $Z'$ in the $\{m_{Z'},g_B\}$ plane from resonance searches in the decays $Z\gamma$~\cite{ATLAS:2022nmi}, $h\gamma$~\cite{CMS:2018cno},  $WW$~\cite{ATLAS:2019nat} . Constraints from the $hZ$ and $ZZ$ channels~\cite{ATLAS:2019nat, CMS:2018kaz} are not strong enough to appear in the plot. The purple line corresponds to the anomalon mass limit $m_{\text{anom}} < 90$~GeV, excluded from searches by the ALEPH and L3 collaborations~\cite{L3:2001xsz, ALEPH:2002gap}.  We overlay the direct dijet resonance constraints taken from Ref.~\cite{Dobrescu:2021vak} for comparison.}
\label{fig:ZpDerivedConstraints}
\end{figure}

In the following, we derive constraints in the $\{m_{Z'},g_B\}$ plane using present limits from narrow resonance searches in data gathered at the LHC. Specifically, we analyze searches for resonances in decays to $\gamma\gamma$, $h\gamma$, $hZ$, $ZZ$/$WW$, and $Z\gamma$~\cite{ATLAS:2016gzy, CMS:2018cno, CMS:2018kaz, ATLAS:2019nat, ATLAS:2022nmi}.  Simulated events for the ALP particle $a$ and the new gauge boson $Z'$ are generated with MadGraph5\_aMC@NLO (MG5\_aMC)3.4.1~\cite{Alwall:2014hca}. We use the narrow-width approximation together with the branching ratios shown in~\figref{Zp_BR_fixgB} to compare the overall cross section with the respective limit. The resulting constraints in the $\{m_{Z'}, g_B\}$ plane are shown in~\figref{ZpDerivedConstraints}.  We remark that the current $WW$ and $h \gamma$ final states probe unrealistically large baryon number gauge couplings, and instead are shown to indicate their relative strength compared to the other collider and indirect probes.  We also show the most recent update from dijet resonance searches~\cite{Dobrescu:2021vak} for comparison as well as the limit on charged anomalons from the ALEPH and L3 collaborations at LEP~\cite{L3:2001xsz, ALEPH:2002gap}.

The most dominant decay channel is the dijet decay, $Z'\to \bar{q} q$, which has already been reviewed recently in Ref.~\cite{Dobrescu:2021vak}.  As evident from~\figref{Zp_BR_fixgB}, the leading exotic decays are $Z' \to a h$, $Z' \to WW$, and $Z' \to ZZ$, while the Wess-Zumino interaction is probed via the $Z' \to \gamma Z$ decay.  From the collider perspective, the relatively suppression of the $Z' \to \gamma Z$ decay in comparison to the others is overcome by the enhanced efficiency for signal photon and leptons and small backgrounds, making the $\gamma Z$ channel the dominant probe of the exotic $Z'$ decays.

The limit of the constraints weakens with increasing $Z'$ mass for two reasons.  First, the production cross sections fall faster than the continuum backgrounds for larger $Z'$ masses.
Second, the model specific branching ratios in~\figref{Zp_BR_fixgB} decrease for masses higher than $2m_t$ because of the dominating $t \bar{t}$ decay channel.  In practice, the $t\bar{t}$ resonance search is not competitive with the dijet resonance search since the invariant mass resolution is diluted because of the presence of neutrinos, unless a boosted, fully hadronic analysis is performed whereby the minimum resonance sensitivity begins at roughly 1~TeV.

\subsection{Constraints on ALPs}
\label{subsec:ALPcollider}

\begin{figure}[ht]
  \centering
    \includegraphics[width=1\textwidth]{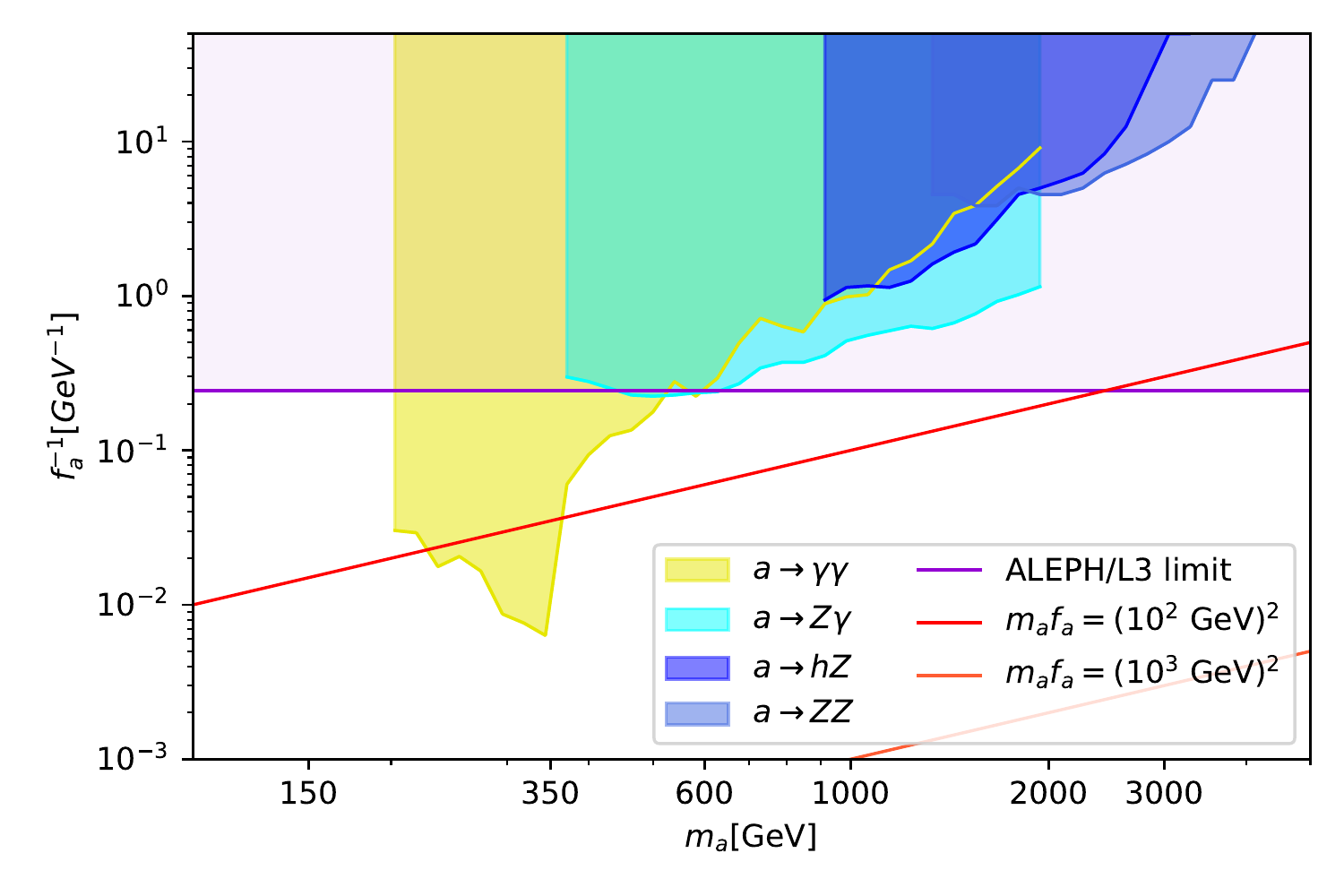}
  \caption{Constraints on the ALP $a$ in the $\{m_a,G_{a\gamma\gamma}\}$ plane from resonance searches in the decays $\gamma\gamma$~\cite{ATLAS:2016gzy}, $Z\gamma$~\cite{ATLAS:2022nmi}, $ZZ$~\cite{ATLAS:2019nat},  and $hZ$~\cite{CMS:2018kaz}. The purple line corresponds to the anomalon mass limit from ALEPH and L3 collaborations~\cite{L3:2001xsz, ALEPH:2002gap} 
  interpreted as a constraint on $f_a$ following~\eqnref{anom_masses} and~\eqnref{vprime_to_fa}.}
\label{fig:ALPDerivedConstraints}
\end{figure}

In this section, we discuss the constraints on ALPs in the $\{m_a, G_{a\gamma\gamma}\}$ parameter space, where new constraints from narrow resonance searches are presented from colliders.  We first discuss the overall status of the diphoton coupling constraint and its interpretation in our model context.  Afterwards we focus on the resulting constraints from narrow resonance searches at the LHC, emphasized separately in~\figref{ALPDerivedConstraints}, provided by various production modes in our model.

\begin{figure}[htb!] 
  \centering
  \includegraphics[width=1.0\textwidth]{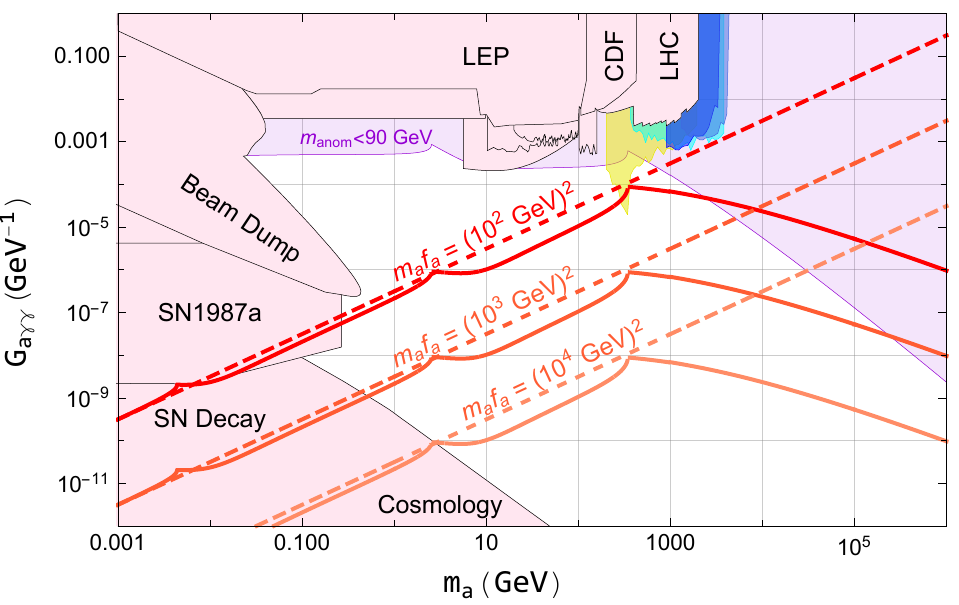}
  \caption{Constraints in the $\{m_a, G_{a\gamma\gamma}\}$-plane for ALP masses $m_a > 1$~MeV for the anomalon mixing angle $\alpha_E = \pi / 4$. The predicted ALP-diphoton couplings are shown for $\sqrt{\chi} \in \{ (10^2 \text{ GeV})^2,\ (10^3 \text{ GeV})^2,\ (10^4 \text{ GeV})^2 \}$ as solid lines with threshold corrections from fermion masses.  For comparison, we illustrate the naive expectation if threshold effects were ignored as dashed lines.  The purple shaded region corresponds to $m_{\text{anom}} < 90$~GeV, excluded from searches by the ALEPH and L3 collaborations~\cite{L3:2001xsz, ALEPH:2002gap}. The updated constraints on exotic ALP signatures at colliders are covered in detail in~\figref{ALPDerivedConstraints}.  The canonical experimental and cosmological bounds are obtained from Refs.~\cite{Jaeckel:2010ni, Alekhin:2015byh, Jaeckel:2015jla, Redondo:2008en, Cadamuro:2011fd, Proceedings:2012ulb, Jaeckel:2012yz, Mimasu:2014nea, Payez:2014xsa, Jaeckel:2017tud, CAST:2017uph}.}
  \label{fig:ax_coll}
\end{figure}

Since we focus on the parameter space where the ALP mass $m_a > 1$~MeV for which the canonical QCD axion is disfavored, we must keep the SM fermions and anomalons dynamical in loop functions.  In particular, recent literature has emphasized the collider scale in context of high-quality axions, where the axion is buffered from UV effects and the topological susceptibility is enlarged~\cite{Kim:1984pt, Choi:1985cb, Randall:1992ut, Rubakov:1997vp, Choi:2003wr, Svrcek:2006yi, Kivel:2022emq}.  The expression for $G_{a\gamma\gamma}$, defined previously in~\eqnref{axL_B}, is 
\begin{equation}
G_{a\gamma\gamma} = \sum_{f\in \text{SM},E_{1,2}} \frac{2e^2}{\pi^2 f_a } N_f Q_f^2 X_A^f \frac{m_f^2}{m_a^2} \ln(\frac{2m_f^2 - m_a^2 + \sqrt{m_a^4 - 4m_a^2 m_f^2}}{2m_f^2})^2 \ ,
\label{eqn:Gagammagamma}
\end{equation}
where axial PQ charges of the anomalons $X_A^{E_{1,2}} = \pm \cos(2\alpha_E) X_B / 2$ were defined in~\eqnref{X_rotation}. The diphoton contribution coming from mass mixings with QCD mesons is suppressed by $\Lambda_\text{QCD}^4/\chi$ and is therefore neglected.

We emphasize that considering the full contribution from SM fermions as well as from anomalons provides a robust prediction for the diphoton coupling of the axion at larger scales.  The corresponding lines are shown in the $\{m_a,G_{a\gamma\gamma}\}$-plane in~\figref{ax_coll} for the topological susceptibilities  $\sqrt{\chi} \in \{ (10^2 \text{ GeV})^2,\ (10^3 \text{ GeV})^2,\ (10^4 \text{ GeV})^2 \}$. The dashed lines correspond to the lines without threshold effects which clearly deviate for ALP masses above the bottom quark threshold and strongly deviate above the top threshold.

The purple shaded region corresponds to searches for charged heavy leptons by the ALEPH and L3 collaborations, which exclude masses $m_\text{anom} < 90$~GeV~\cite{L3:2001xsz, ALEPH:2002gap}.  This implies a lower bound on $f_a \gtrsim 5$~GeV which we can derive from the mass of the lighter electron anomalon given by~\eqnref{anom_masses} and the relationship between $v'$ and $f_a$ from~\eqnref{vprime_to_fa}, using Yukawa couplings of $4\pi / 3$ and anomalon mixing angles of $\beta = 0.05$ and $\beta' = \pi/4$.  For this fixed $f_a$ bound, we use~\eqnref{Gagammagamma} to extract an upper bound on $G_{a\gamma\gamma}$, shown as the purple line labeled $m_{\text{anom}} < 90$~GeV.  We remark that this line, for $m_a \lesssim 2 m_t$, is roughly independent of $m_a$ since the fixed $f_a$ can be considered independent of varying $m_a$ by rescaling the topological susceptibility.  The increased sensitivity in $G_{a \gamma \gamma}$ from the anomalon bound above $m_a \gtrsim 2 m_t$ arises from the logarithmic dependence on $m_a$ for light fermion masses in~\eqnref{Gagammagamma}.

We now focus on the collider searches for ALPs with $m_a > 100$~GeV, as shown in~\figref{ALPDerivedConstraints}.  Considering the full contribution for the diphoton coupling in this model enhances previously set constraints that were established for a general ALP. We simulate the ALP cross section from gluon fusion using MadGraph 5~\cite{Alwall:2014hca} as before and apply the full expressions for the branching ratios into the resonance search channels $\gamma \gamma$, $hZ$, $ZZ$, and $Z \gamma$~\cite{ATLAS:2016gzy, CMS:2018kaz, ATLAS:2019nat, ATLAS:2022nmi}.  Although the DFSZ-type ALP typically has a dominant decay into $t\bar{t}$ above the top threshold, interpreting the collider constraints is highly non-trivial because of large interference effects between the continuum $t\bar{t}$ production from gluons and the ALP signal that can wash away the expected resonance~\cite{Carena:2016npr}.  We leave the derivation of ALP constraints in the $t\bar{t}$ final state for future work.

For the ALP, the $\gamma\gamma$ resonance searches are the most constraining, since other channels are more suppressed and also have weaker collider sensitivity from larger backgrounds.  We emphasize that considering the full contribution for the diphoton coupling leads to a significant shift at the top threshold that is a consequence of the ALP coupling to the top quark.  As a result, we see that many of the exotic decays for the ALP can be competitive and exceed the sensitivity from the diphoton channel, and we expect that further gains in sensitivity will come when these searches are analyzed with available luminosity.

\section{Conclusions}
\label{sec:conclusions}

In this work, we have studied a DFSZ-type axion model in an gauged baryon-number extension of the Standard Model.  Our aim was to establish the patterns of axion and $Z'$ effective couplings when a canonical anomalous global symmetry of the quarks, namely baryon number, is promoted to a gauge symmetry, in the particular case when the quarks also carry the Peccei-Quinn symmetry.

We calculated the Wilson coefficients which arise at low energies in a general way accounting for possible flavor-violating interactions as well as for $U(1)$ basis transformations of the interacting fermions. For this purpose we set up a general axion interaction basis which contains operators which do not appear in the flavor conserving limit, like the axion coupling to a gauge boson and the commutator current $J^{\mu,a}_{[ I , \text{PQ}] }$.  We found that this operator also appears in the standard DFSZ model for the interactions to the $W$ bosons, reflecting the flavor-changing effects from the CKM matrix.  Some new Wilson coefficients in our EFT include interactions between the axion, $Z'$ and $\gamma$ and the axion, $Z'$ and Higgs boson, which provide novel decay modes for both the axion or the $Z'$ boson, depending on the mass hierarchy.  Our charged anomalons also characteristically canceled in their contributions to the axion-diphoton coupling, with the residual coupling only driven by the SM fermions.

We then presented the phenomenology of the model, encompassing both the new decay channels of the axion and the $Z'$ boson from the effective description.  We showed the ALP can have a relatively large branching fraction to $hZ$ nearing 10\%, while the dominant decay above the top threshold is to $t\bar{t}$, which requires a special analysis given the non-trivial interference with the continuum top pair production which we leave for future work.

Present limits from narrow resonance searches at the LHC are derived and presented in the $\{m_{Z'},g_B\}$ plane in~\figref{ZpDerivedConstraints} and in the $\{m_a,G_{a\gamma\gamma}\}$ plane in \figref{ax_coll}.  Our results systematically capture all of the diverse signals arising from the EFT description and demonstrate complementarity between the different search channels at the ATLAS and CMS experiments, especially those beyond the $\gamma \gamma$ coupling typically studied for axions and ALPs.

\section*{Acknowledgments}
\label{sec:acknowledgments}

This research is supported by the Cluster of Excellence PRISMA$^+$, ``Precision Physics, Fundamental Interactions and Structure of Matter" (EXC 2118/1) within the German Excellence Strategy (project ID 39083149).  FY would like to express special thanks to the Mainz Institute for Theoretical Physics (MITP) of the Cluster of Excellence PRISMA+ for its hospitality and support.


\bibliographystyle{apsrev4-1}
\bibliography{quotations}

\end{document}